\documentclass{jaa}
\usepackage{natbib}
\usepackage{graphicx}
\usepackage{aas_macros}
\usepackage{amsmath} 
\usepackage{txfonts}
\usepackage{balance}
\usepackage{float}
\usepackage{multirow}
\DeclareUnicodeCharacter{2212}{-}
\usepackage{hyperref}
\usepackage{float}  
\begin{document}\sloppy

\title{Thermal and Turbulence Characteristics of Fast and Slow Coronal Mass Ejections at 1 AU}

\author{Soumyaranjan Khuntia \orcid{0009-0006-3209-658X}\textsuperscript{1,2,*}, and Wageesh Mishra \orcid{0000-0003-2740-2280}\textsuperscript{1, 2}}
\affilOne{\textsuperscript{1}Indian Institute of Astrophysics, II Block, Koramangala, Bengaluru 560034, India\\}
\affilTwo{\textsuperscript{2} Pondicherry University, R.V. Nagar, Kalapet 605014, Puducherry, India}

\twocolumn[{

\maketitle

\corres{soumyaranjan.khuntia@iiap.res.in}

\begin{abstract}
Understanding the thermal and turbulence properties of interplanetary coronal mass ejections (ICMEs) is essential for analyzing their evolution and interactions with the surrounding medium. This study explores these characteristics across different regions of two distinct ICMEs observed at 1 AU, utilizing in-situ measurements from the Wind spacecraft. The polytropic indices (\(\Gamma_e\) for electrons and \(\Gamma_p\) for protons) reveal significant deviations from adiabatic expansion, suggesting sustained heating mechanisms within the ICMEs even at 1AU. The effective polytropic index ($\Gamma_{\text{eff}}$) of the magnetic ejecta (ME) in both ICME1 and ICME2 is found to be near-isothermal ($\Gamma_{\text{eff}} = 0.88$ and $0.76$), aligning with measurements near the Sun, highlighting consistent heating across heliospheric distances. Spectral analysis at the inertial scale reveals Kolmogorov-like turbulence in the fast ICME1's ME, while the ME of the slower ICME2 exhibits less developed turbulence with a shallower spectral index (\(\alpha_B\)). The turbulence analysis in the dissipation scale indicates that the ME of slower ICME2 is less affected by the ambient medium than the faster ICME2. The MEs of both ICMEs show magnetic compressibility much smaller than unity (\(C_B<1\)), suggesting dominant Alfvénic fluctuations in the MEs. Notably, the partial variance of increments (PVI) method identifies more intermittent structures, such as current sheets and reconnection sites, in the sheath and post-ICME regions. Higher PVI values correlate with regions of increased electron and proton temperature (for the sheath region), as well as higher \(C_B\) values, highlighting their role in local energy dissipation. These results underscore the importance of ongoing heating and turbulence processes in shaping the evolution of ICMEs. 

\end{abstract}


}]

\doinum{12.3456/s78910-011-012-3}
\artcitid{\#\#\#\#}
\volnum{000}
\year{2025}
\pgrange{1--}
\setcounter{page}{1}
\lp{1}

\section{Introduction}

Coronal mass ejections (CMEs) are large-scale expulsions of magnetized plasma from the solar corona, and they are historically termed interplanetary CMEs (ICMEs) away from the Sun \citep{Webb2012,Temmer2023}. Upon reaching Earth, ICMEs can trigger severe geomagnetic storms, posing significant risks to both space-borne and terrestrial technologies \citep{Gosling1993, Baker2009, Temmer2021}. A comprehensive understanding of their thermal evolution, a complex interplay of plasma heating, cooling, and energy dissipation processes, is fundamental to improving space weather prediction models and better understanding the Sun-Earth connection.

The thermal state of an ICME evolves as it propagates outward from the Sun in the solar wind, governed by complex interactions within and with the ambient medium. The heating of CMEs has also been reported through spectroscopic observations of the erupting material \citep{Filippov2002, Lee2017, Reva2023} and by analyzing ionization states at 1 AU \citep{Rakowski2007,Lepri2012}. The Flux Rope Internal State (FRIS) model with 3D kinematics as inputs has also been used to understand the CMEs thermodynamics in coronagraph and heliospheric imaging field of view \citep{Mishra2020, Mishra2023err, Khuntia2023}. Recent studies using the FRIS model found that both fast and slow CMEs tend to evolve toward an isothermal state beyond a height of approximately 3–7  $R_\odot$ \citep{Khuntia2023,Khuntia2024}. Studies on the thermodynamic evolution of ICMEs within their different substructures using in situ observations are limited. The thermal state of ICME plasma can be characterized by the polytropic index ($\Gamma$), assuming the plasma evolves through a polytropic process. The polytropic equation, given by $T n^{1-\Gamma} = \text{constant}$, describes the relationship between plasma density ($n$) and temperature ($T$). Without involving complex energy equations, the $\Gamma$ serves as a key parameter to determine the thermodynamic behavior of the system under a polytropic approximation. Although the polytropic equation is derived for an ideal neutral gas undergoing slow, quasi-static processes, it has proven useful for characterizing key features of various space plasmas such as the solar corona, solar wind, and ICMEs, where the plasma is magnetized and weakly collisional/collisionless \citep[e.g.,][]{Wu1999a, Linker2003, Hayashi2006, Riley2006, Hu2008}. In such environments, direct application of the full energy equation is often infeasible due to the complexity of unresolved microphysical processes like turbulent dissipation, heat conduction, wave-particle interactions, and intermittent heating. While not giving a perfect physical description of the process, the polytropic relation offers an empirical framework to explain the net effect of these processes on the plasma's thermal state. A value of $\Gamma = 5/3$ corresponds to adiabatic expansion with no heat exchange, whereas $\Gamma < 5/3$ indicates the presence of additional heating processes during plasma evolution. Previous studies have reported that the $\Gamma$ value for ICMEs at large heliocentric distances typically ranges between 1.15 and 1.33 \citep{Liu2005,Liu2006}. This suggests significant local plasma heating within the ICME, even far beyond 1 AU, likely driven by small-scale processes such as magnetic reconnection, turbulence, or interactions with the solar wind. A recent study on the great geomagnetic storm, employing an analytical model combined with the polytropic approach, found consistent plasma heating across multiple heliocentric distances \citep{Khuntia2025}.

Different physical processes responsible for CME heating can lead to spatial inhomogeneities in thermal properties across various regions of a single ejecta. Since in-situ observations provide time-series data along the spacecraft trajectory, analyzing smaller time intervals effectively samples distinct spatial segments of the CME. This approach offers better insights into localized thermal states, as opposed to fitting the entire duration, which may obscure regional variations and yield poor correlations. Previous works have employed this interval-based approach to study solar wind \citep{Nicolaou2014, Nicolaou2020} and ICME structures at 1 AU \citep{Dayeh2022}, revealing detailed thermal behavior. \citet{Dayeh2022} suggest that the polytropic index varies across different regions of an ICME, such as the sheath, the magnetic ejecta (ME), and the ambient solar wind, reflecting diverse thermal states in each region. There are limited studies that understand the turbulence within ICMEs, which could play a vital role in energy dissipation and plasma heating across various spatial and temporal scales within ICMEs. While ambient solar wind turbulence has been extensively studied, the turbulent characteristics within ICMEs and their evolution remain active research areas.

It is understood that energy injected at large scales cascades through inertial scales before dissipating at smaller kinetic scales, such as ion and electron gyroscales \citep{Matthaeus1999, Matthaeus2021}. The power spectral density (PSD) of these turbulent magnetic fluctuations follows distinct power-law scaling in the frequency ($f$), with injection scales showing $f^{-1}$, inertial ranges exhibiting slopes between $f^{-5/3}$ and $f^{-3/2}$, and dissipation scales ranging between $f^{-2}$ and $f^{-4}$ \citep{Kraichnan1965, Bruno2013}. The slow solar wind typically shows greater variability and steeper spectral slopes in the inertial range than the fast solar wind, often approaching the Kolmogorov index ($-5/3$) at 1 AU \citep{Smith2006, Chen2020}. The kinetic-range spectral index for solar wind turbulence is usually less than $-2.8$, implying an ongoing turbulent energy cascade \citep{Sahraoui2009}. These spectral characteristics are influenced by compressibility, anisotropy, and intermittent structures, which are also observed in ICME turbulence \citep{Kilpua2020}. ICMEs often consist of a shock, sheath, and magnetic ejecta (ME). A subset of ME is the magnetic flux rope (MFR), which shows a strong magnetic field, low plasma beta ($\beta \ll 1$), and reduced density and temperature in the in-situ measurements \citep{Zurbuchen2006}. Turbulence within ICME regions differs significantly from that of the surrounding solar wind. Turbulence is especially prominent in the sheath region of ICMEs, where compressed solar wind particles and shock interactions lead to enhanced levels of fluctuations. Conversely, the ME region exhibits more structured turbulence, characterized by smoother magnetic field variations and coherent large-scale features such as MFR, compared to the highly irregular fluctuation in the sheath. Despite these differences, both regions may contribute to plasma heating through distinct turbulent processes. Recent studies suggest that turbulence in the ICME sheath at 1 AU deviates from classic power-law indices of $-5/3$ (Kolmogorov) and $-3/2$ (Kraichnan), indicative of evolving turbulence \citep{Kilpua2020}. Turbulence in the sheath is likely not fully developed, due to the ongoing influence of dynamic processes such as shock interactions and compressive flows near the ejecta’s leading edge \citep{Kilpua2020}. Moreover, these deviations can be attributed to the breakdown of the assumptions of homogeneity and isotropy, as the presence of large-scale magnetic field fluctuation, shocks, and flow shear in the sheath region alters the energy cascade and results in a more complex spectral behavior. In contrast, statistical study of the turbulence spectra within the MFR shows average spectral indices of -1.56 and -2.01 in the inertial and dissipation range, respectively \citep{Hamilton2008}. Recent work by \citet{Good2023} using Solar Orbiter and Parker Solar Probe data revealed that ICME MFR spectral indices often remain close to $-5/3$ at all the observed radial distances. Understanding the interplay between thermal processes and turbulence in CME regions is crucial for unraveling how energy is transferred, cascaded, and ultimately dissipated in ICME plasma during its interplanetary propagation.

The present study extends our previous work \citep{Khuntia2023} on the analysis of the thermodynamic properties of a fast (2011 September 24) and a slow CME (2018 August 20) near the Sun, within a height range of 3–20 $R_\odot$, using a combination of remote observations and analytical modeling. These two geoeffective CMEs are identified in the in situ observations near 1 AU as ICME1 (26 September 20211) and ICME2 (25 August 2018) \citep{Wood2016,Gopalswamy2022}. In the present study, we focus on investigating the thermal state and turbulence properties of the corresponding ICMEs at 1 AU using in-situ observation. We analyzed the variation of the thermal state for both electron and proton by estimating the electron and proton polytropic index across different substructures within the ICME. By analyzing the polytropic index and magnetic field turbulence characteristics across different ICME regions, we aim to explore thermal and turbulence properties for both slow and fast ICMEs.

\section{Data Overview and Methodology }

\subsection{Deriving polytropic index (\texorpdfstring{$\Gamma$}{Gamma})}

We utilized high-resolution data from the Wind spacecraft to analyze the two selected ICMEs, ICME1 (26 September 2011) and ICME2 (25 August 2018). Our focus is on the highest available resolution data for both electrons and protons to capture finer variations in the derived thermal states. The 9-second resolution data of electron number density $n_e$ and temperature $T_e$ from the Wind/Solar Wind Experiment (SWE; \citealt{Ogilvie1995}) instrument are used to derive the electron polytropic index ($\Gamma_e$), while 92-second resolution proton number density $n_p$ and temperature $T_p$ data were obtained from the Wind/SWE to derived proton polytropic index ($\Gamma_p$). All data were retrieved from the \textit{Coordinated Data Analysis Web (CDAWeb)}. The higher resolution of the electron data captures finer variations in the thermal states within a localized region compared to the protons. However, this resolution difference does not affect our analysis of the mean thermal state for both electrons and protons. We will examine the electron and proton polytropic index differences as earlier studies have suggested that electron temperature in ICMEs is higher than proton temperature \citep{Zurbuchen2006}. In addition to the ICME structures, 3 hours of pre and post-ICME solar wind is also analyzed to compare the turbulence and thermodynamic properties of the ICME with the surrounding medium.


\begin{figure}[ht]
    \centering
    \includegraphics[width=0.95\linewidth,trim={0pt 5pt 0pt 10pt}, clip]{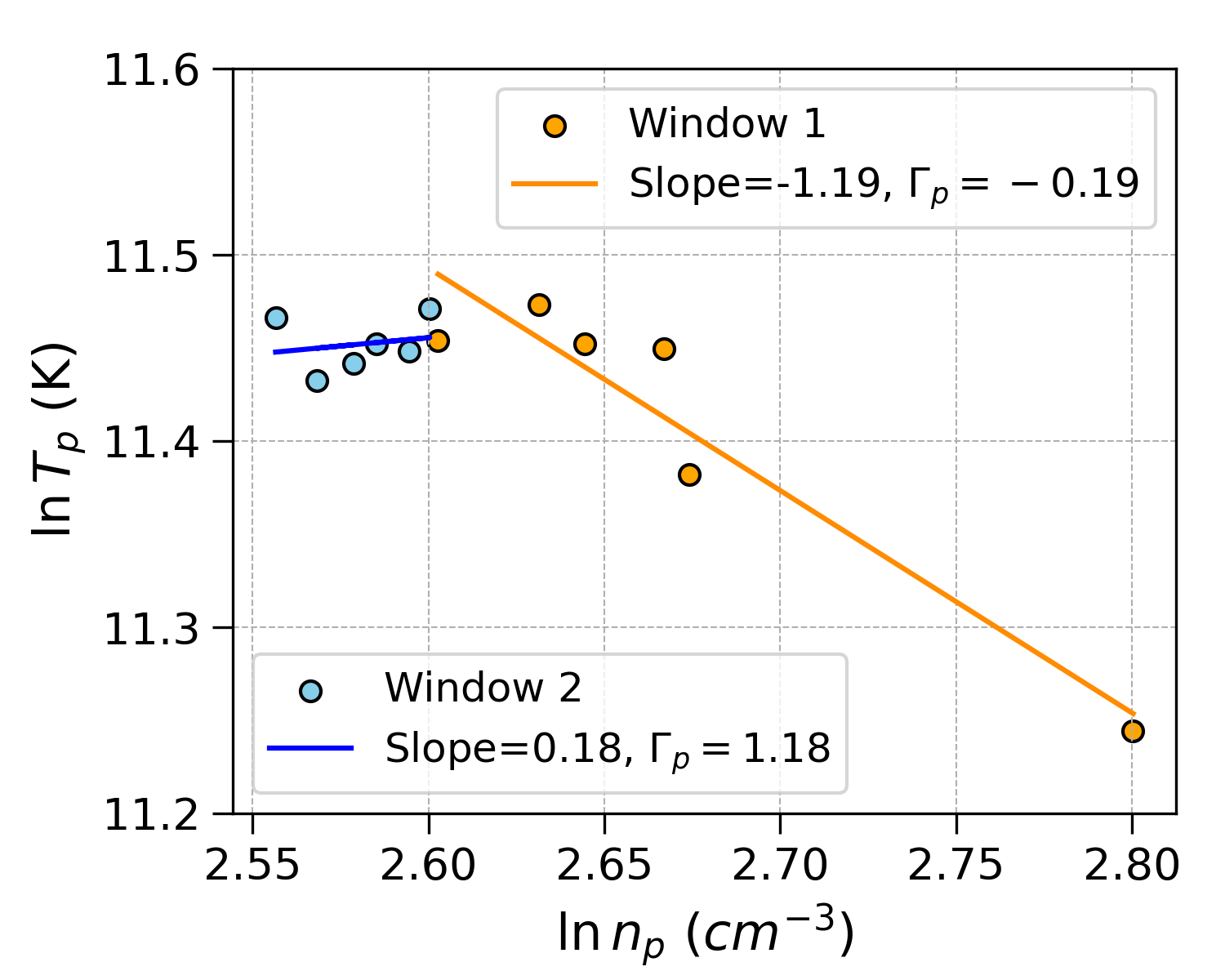}
    \caption{ Measurement of proton polytropic index ($\Gamma_p$) for two random windows with 6 data points}
    \label{fig:demo_gamma}
\end{figure}


In our study, the polytropic index was calculated using a linear fit between $\log T$ and $\log n$, along with the corresponding Pearson correlation coefficient (CC) and p-value (p). To enhance the robustness of $\Gamma$ estimates, we applied this fitting procedure to moving sub-intervals with a step size of 9 and 92 seconds (i.e., resolution of data) for electron and proton data, respectively. We further filtered the sub-intervals based on the quality of the log-log linear fit, primarily relying on $\Gamma$ values corresponding to CC $>$ 0.8 and p $<$ 0.05. We selected an optimal sub-interval duration of 6 data points to fit density and temperature variations, estimating the $\Gamma$ for each sub-interval (Figure \ref{fig:demo_gamma}). This choice was based on a combination of statistical and physical considerations. Specifically, we tested sub-interval durations ranging from 3 to 10 data points for both proton and electron datasets and found that using 6 data points yielded the maximum number of reliable $\Gamma$ estimates that satisfied our correlation criteria. Longer sub-intervals were excluded, as they are more likely to encompass multiple plasma streamlines or evolving structures, which can obscure localized thermal signatures and reduce the correlation between density and temperature. This filtering ensured that the computed $\Gamma$ values most effectively captured the thermal state of each localized region within the ejecta.

\begin{figure}[ht]
    \centering
    \includegraphics[width=\linewidth]{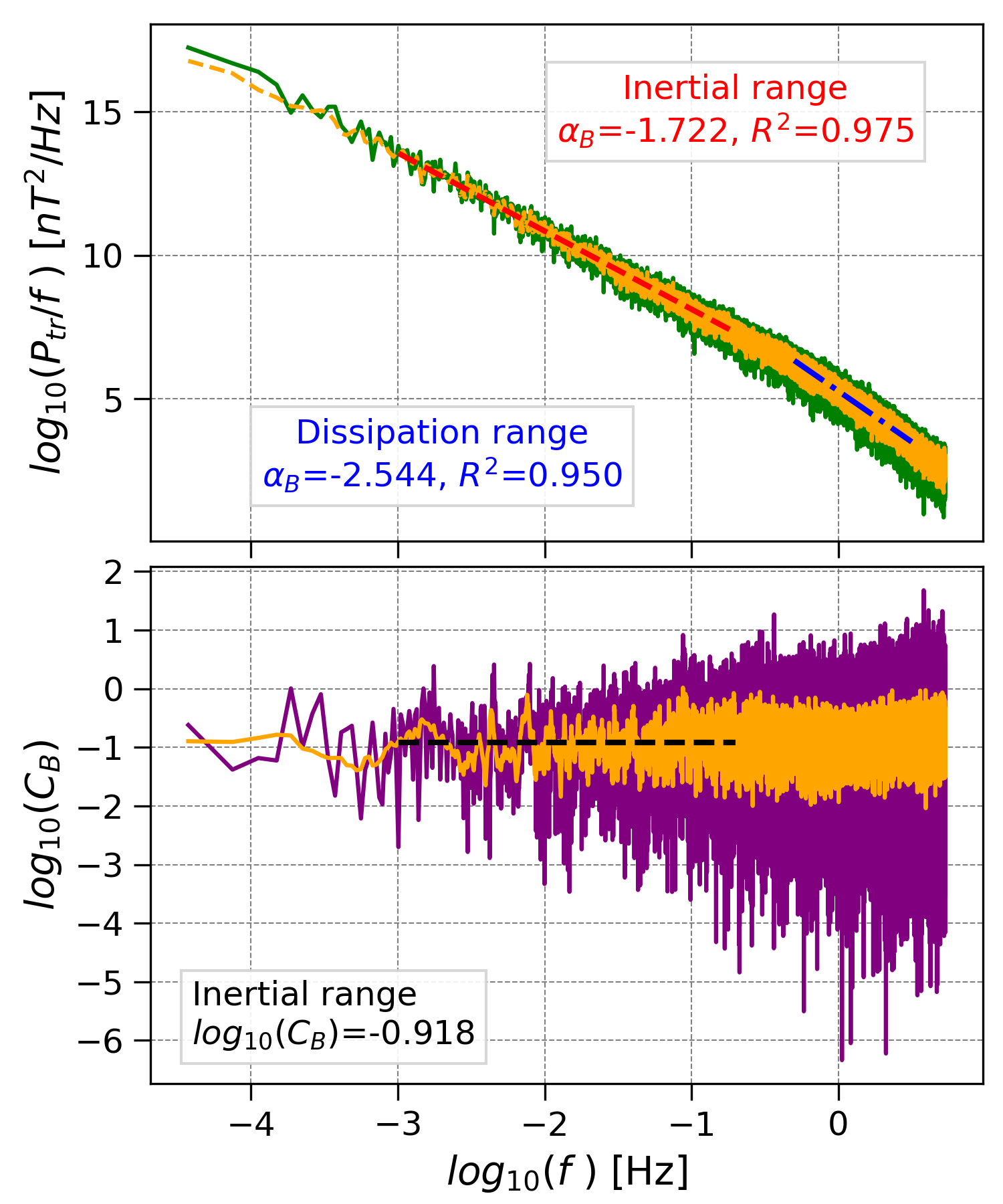}
    \caption{ (a) Variation of trace magnetic power spectra per frequency and (b) magnetic compressibility $C_B$ with frequency for the sheath region of the ICME1. The yellow curve on (a) represents the 3-point sliding average values of power spectra. The red and blue dash lines were fitted to the inertial ($-3 < \log_{10} f < -0.7 $) and dissipation ($-0.3 < \log_{10} f < 0.5 $) scale of 3-point sliding average values of trace magnetic PSD. The yellow curve on (b) panels represents 10-point moving average values of $C_B$. The black dashed line shows the average value of $C_B$ over the inertial range. }
    \label{fig:fast_spec_sheath}
\end{figure}

\subsection{Deriving turbulence characteristics}

The 11 Hz (0.092 seconds) resolution magnetic field measurements from the Magnetic Field Investigation (MFI; \citet{Lepping1995}) are used for turbulence analysis of the selected ICMEs. This High-resolution data ensures that the fine-scale turbulent structures are adequately resolved, allowing for accurate spectral analysis. The time (t) series magnetic field fluctuations are computed as, $\delta B(t) = B(t) - \langle B(t) \rangle \quad \text{and} \quad \delta B_i(t) = B_i(t) - \langle B_i(t) \rangle,$ where \(i = (x, y, z)\), and \(\langle ... \rangle\) represents the average value over the selected regions that will be discussed in the next section. We used Fast Fourier Transformation (FFT) to convert the time-series fluctuations 
\(\delta B_i(t)\) into their frequency-domain counterparts \(\delta B_i(f)\) to analyze the turbulence spectra of these magnetic field fluctuations. The Fourier-transformed magnetic field components are \(\delta B_x(f)\), \(\delta B_y(f)\), and \(\delta B_z(f)\), and their respective power spectra are represented as \[ P_x = |\delta B_x(f)|^2, \quad P_y = |\delta B_y(f)|^2, \quad P_z = |\delta B_z(f)|^2.\] The total and trace magnetic field power spectral density (PSD) are given, respectively, by \[P_t = |\delta B(f)|^2, \quad \text{and} \quad P_{tr} = P_x + P_y + P_z \] where \(f\) is the frequency components of the magnetic field fluctuations in the time series data. PSD represents how the power (or energy) of magnetic field fluctuations is distributed across different frequencies. The turbulence characteristics are described by a power-law dependence of the trace magnetic field PSD on frequency, expressed as $ P_{tr} \sim f^{\alpha_B}$, where $\alpha_B$ is the spectral slope and determines how the energy in the turbulence decays with frequency. We applied a sliding three-point average to smooth the PSD spectrum before performing a linear fit to it in log-log space to estimate $\alpha_B$, which is the slope of the fitting. This smoothing helps reduce noise and improve the accuracy of the spectral slope estimation. The spectral slope provides insights into the nature of turbulence within the ICME plasma.

 To better understand the dominant nature of magnetic field fluctuations and the sources from where they arise, one can calculate the magnetic compressibility ($C_B$), which is the ratio of total and trace magnetic field PSD, $C_B = P_{t}/P_{tr}$ \citep{Telloni2021}. It gives a quantitative measure of the Alfvénic-dominated incompressible fluctuations within ICME plasma over compressible fluctuations from slow/fast-mode magnetosonic waves. Figures \ref{fig:fast_spec_sheath}(a) and (b) show the variation of trace magnetic PSD ($P_{tr}$) per frequency and magnetic compressibility $C_B$ with frequency, respectively, for the sheath region of the ICME1. The trace magnetic PSD is fitted with two distinct power laws corresponding to two frequency regimes (Figure \ref{fig:fast_spec_sheath}(a)): the inertial scale ($-3 < \log_{10} f < -0.7$) and the dissipation scale ($-0.3 < \log_{10} f < 0.5$). The inertial frequency range is identified as a segment with a nearly straight line and moderate negative slope (typically close to $-5/3$ or $-3/2$). The lower limit of the inertial range is typically marked by the transition from the energy-containing range to turbulent scales. For solar wind and ICME turbulence, the lower limit of the inertial range typically lies around $\log_{10} f < -3$ \citep{Bruno2013, Chen2020, Telloni2021}. The upper limit corresponds to the frequency where the slope begins to steepen, indicating the onset of dissipation processes. The dissipation range follows, where the slope steepens significantly (e.g., to $-2$ or more negative), indicating energy dissipation and smaller-scale kinetic processes. These range transitions appear as distinct slope changes in the plotted PSD curve (Figure \ref{fig:fast_spec_sheath}(a)). The transition point between these two regions is referred to as the spectral break \citep{Markovskii2008, Chen2014sbreak}, which varies depending on the region and is beyond the scope of this study. The average value of $C_B$ is calculated over the inertial scale, shown as a black dashed line in Figure \ref{fig:fast_spec_sheath}(b).

\begin{figure*}[ht]
    \centering
    \includegraphics[width=0.49\linewidth]{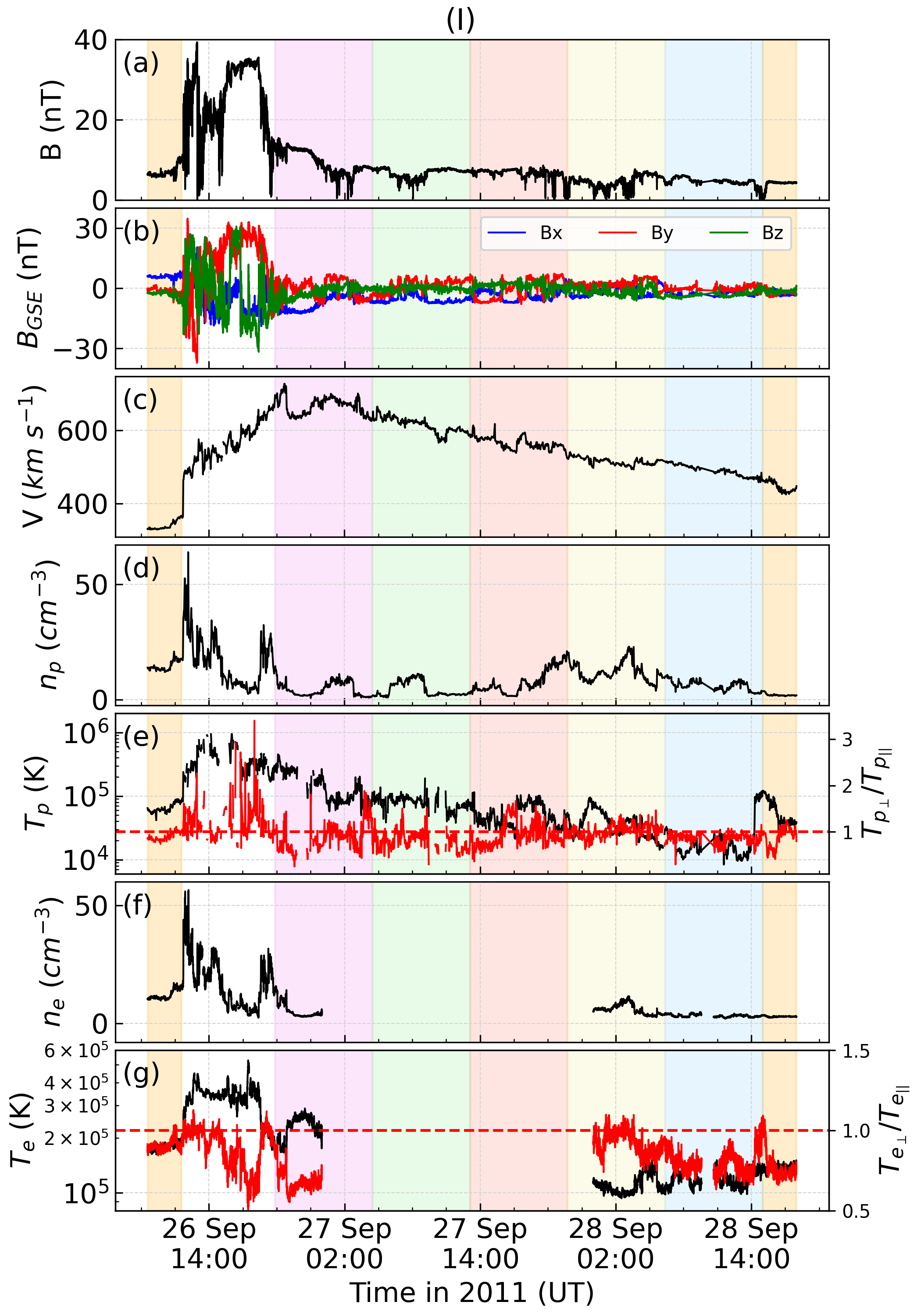} 
    \includegraphics[width=0.49\linewidth]{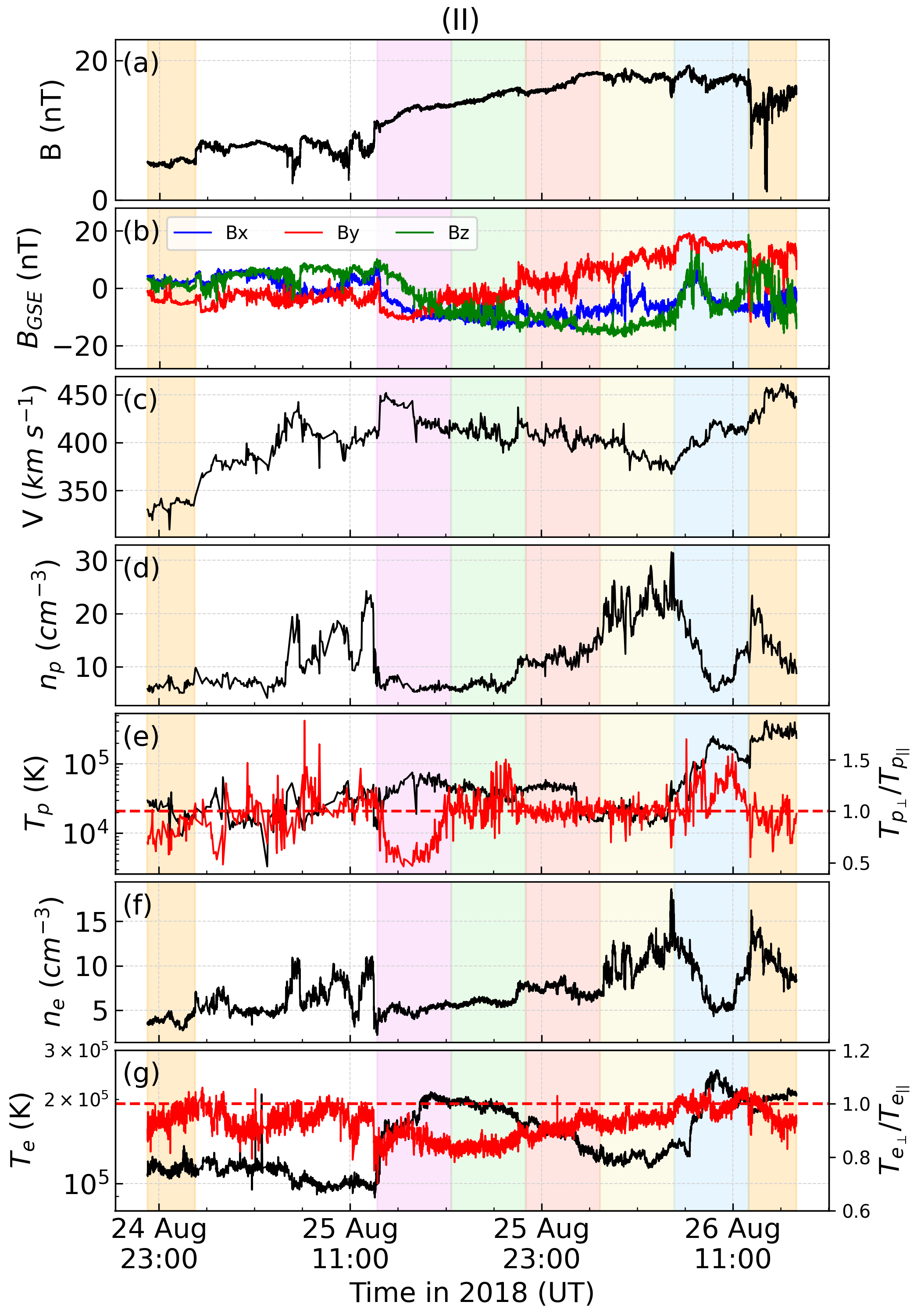}
    \caption{Observed in-situ parameters for (a) ICME1 and (b) ICME2 using magnetic and plasma data from the Wind spacecraft. The orange-shaded regions at the beginning and end represent the pre and post-ICME solar wind. The white-shaded area marks the sheath region, while the multicolored-shaded region between the sheath and post-ICME solar wind corresponds to the magnetic ejecta, divided into five equal parts.}
    \label{fig:insitu_wind}
\end{figure*}

Intermittency in ICMEs is the irregular and highly localized concentration of turbulence and energy dissipation and arises from the concentration of sharp magnetic field gradients/fluctuations within small sub-volumes \citep{NOVIKOV1971}. Identification of intermittency is important because it provides insights into the non-uniform and highly localized nature of energy dissipation in turbulent systems, including plasmas like the solar wind and ICMEs. The Partial Variance of Increments (PVI) method is widely used to study the kinematics and formation of intermittency or coherent structures in space plasmas \citep{Greco2008, Greco2009}. The PVI method is highly sensitive to directional shifts, changes in magnitude, and sharp gradients in the vector magnetic field $B$. The PVI measures the magnitude of the change in a vector magnetic field over a specified time lag, normalized by its variance. The method consists of defining a series in terms of the vector magnetic field increment along a linear trajectory in time (t) labeled by \[ \Delta B(t,\tau)= B(t+\tau)-B(t)\] where $\tau$ is the separation time or time lag. The PVI can be defined via the normalized quantity, \[ PVI = \frac{|\Delta B(t, \tau)|}{\sqrt{\langle |\Delta B(t, \tau)|^2 \rangle}} \] where \(\langle ... \rangle\) represents an average calculated over a sufficiently large trailing sample from the dataset to ensure statistical significance. \citet{Servidio2011} showed that longer averaging periods dilute large signal variations with smaller ones, amplifying PVI peaks. In contrast, shorter periods result in weaker, less bursty signals due to larger denominators in intervals with significant variations. Thus, averaging intervals should not be too short, with those near the correlation length providing stable averages in homogeneous turbulence. For this study, we calculate the average over the whole data \citep{Emiliya2021}, including pre-ICME, sheath, ME, and post-ICME regions. Also, we choose three different $\tau$ values, such as 0.18, 9.2, and 92 seconds, to identify the intermittency in both dissipation and inertial scales \citep{Kilpua2020}.

\section{Results and discussion}
\subsection{In-situ observation overview  }

\begin{table*}[ht]
\centering
\caption{Mean values of the observed in-situ parameters of various regions associated with ICME1 and ICME2 using magnetic and plasma data from Wind spacecraft.}
\label{tab:cme_table}
\begin{tabular}{|c|cc|cc|cc|cc|cc|}
\hline
\multirow{3}{*}{Parameters} & \multicolumn{8}{c|}{Mean values in various regions of ICME1 and ICME2} \\  
\cline{2-9}
& \multicolumn{2}{c|}{Pre-ICME} & \multicolumn{2}{c|}{Sheath} & \multicolumn{2}{c|}{ME} & \multicolumn{2}{c|}{Post-ICME} \\  
\cline{2-9}
& ICME1 & ICME2 & ICME1 & ICME2 & ICME1 & ICME2 & ICME1 & ICME2 \\
\hline
$B$ [$nT$] & 6.8 & 5.3 & 25.7 & 7.7 & 6.5 & 15.9 & 3.5 & 14.1 \\
\hline
$T_p$ [$10^4 K$] & 6.4 & 2.0 & 44.1 & 2.4 & 7.2 & 5.5 & 7.7 & 26.5 \\
\hline
 $n_p$  [$cm^{-3}$] & 14.5 & 6.3 & 16.9 & 10.3 & 7.3 & 12.0 & 2.3 & 13.7 \\
\hline
$T_e$  [$10^4 K$] & 17.9 & 11.3 & 32.3 & 10.9 & 14.2 & 16.5 & 13.6 & 19.9 \\
\hline
$n_e$ [$cm^{-3}$]  & 12.0 & 3.8 & 17.5 & 6.3 & 4.6 & 7.5 & 2.8 & 10.6 \\
\hline
$v$ [$km$ $s^{-1}$]   & 336 & 334 & 558 & 396 & 578 & 406 & 454 & 443 \\
\hline
$v_{exp}$ [$km$ $s^{-1}$]   & - & - & - & - & 186 & 35 & - & - \\
\hline
\end{tabular}
\end{table*}

Figure \ref{fig:insitu_wind} (I) and (II) illustrate the two ICME events on 26 September 2011 (ICME1) and 25 August 2018 (ICME2), respectively, observed by the Wind spacecraft at 1 AU. The ICME1 shock onset, marked by the starting of the white-shaded region at 11:34 UT on 26 September 2011, is identified by sudden enhancements in the total magnetic field ($B$), proton bulk velocity ($V$), proton number density ($n_p$), and proton temperature ($T_p$). The white shade region represents the ICME sheath in both ICME1 and ICME2. This sheath region is characterized by enhanced $B$ and fluctuating $B_x$, $B_y$, and $B_z$, high $V$, as well as elevated and varying plasma parameters. Beyond the sheath region lies the magnetic ejecta (ME) of the ICME. For our study, we divided the ME into five distinct parts, represented by multi-colored shaded regions. We divided the magnetic ejecta (ME) into five parts to better understand its structure and the changes across different regions. It may be possible that Part 1 has some influence from the adjacent sheath region, while Part 5 may have some influence from the adjacent post-ICME solar wind. Part 3, situated at the probable center of the ME, primarily represents its inherent properties, largely unaffected by external regions. Additionally, Parts 2 and 4 are included to bridge the transitions between these key regions—between Part 1 and Part 3 and between Part 3 and Part 5, respectively. The thermal and turbulence characteristics of each part will shed light on the evolution of CME plasma properties and its connection with the ambient solar wind. We also included 3 hours of pre and post-ICME solar wind data for our analysis, marked by orange-shaded regions for both the ICMEs.

The red color curve in panels (e) and (g) of Figure \ref{fig:insitu_wind} show the temperature anisotropy ($T_{aniso}$=$T_{\perp}/T_{\parallel}$) for proton and electron, respectively. $T_{\perp} $ and $T_{\parallel}$ is the temperature perpendicular and parallel to the magnetic field direction. Anisotropic heating and cooling processes in ICME and solar wind plasma preferentially affect one temperature component, leading to $T_{\perp} \neq T_{\parallel}$. Thus, temperature anisotropy is known to generate various instabilities and electromagnetic waves in the plasma \citep{Gary1993, Denton1994}. The sheath region of ICME1 has $T_{aniso}>1$ for the proton, suggesting more heating in a perpendicular direction to the magnetic field (Figure \ref{fig:insitu_wind}I(e)), which could arise mirror mode waves and Alfven ion cyclotron waves \citep{Liu2006waves, AlaLahti2019, Verscharen2019}. In contrast, the electrons in the sheath of ICME1 show  $T_{aniso}<1$, which could lead to firehose instabilities and can generate low-frequency whistler waves \citep{Hollweg1970, Hellinger2000}. We also note that $T_{aniso}\approx1$ around the boundary of pre-ICME and sheath, sheath and ME, and ME and post-ICME for ICME1 electrons (Figure \ref{fig:insitu_wind}(I)(g)). Compared to ICME1, ICME2 has an overall less proton temperature anisotropy (Figure \ref{fig:insitu_wind}(II)(e)). The sheath of ICME2 has a mix of both dominate $T_{\perp}$  and $T_{\parallel}$ components. Notably, the first part of ME, near the sheath region, has $T_{aniso}<1$, whereas all other parts of ME have $T_{aniso}>1$. Interestingly, the last two parts of ME show two different characteristics. Part 4 has more density with less temperature, whereas Part 5 has less density and more temperature. The high density and cooler material towards the trailing edge of the ICME2 could be the filament material as this ICME corresponds to the CME-associated filament eruption \citep{Mishra2015b,Gopalswamy2022}. The higher density in Part 4 could also be due to compression from Part 5, which has less density and high-temperature material, possibly related to fast solar wind. fast solar wind streams. Importantly, the post-ICME solar wind is faster than ICME2 and is associated with a co-rotating interaction region (CIR) and followed by a high-speed stream (HSS) region \citep{Mirko2020}.

Table \ref{tab:cme_table} summarizes the mean values of in-situ parameters for various regions associated with ICME1 and ICME2, derived from Wind spacecraft magnetic and plasma data. The sheath region of ICME1 exhibits a significantly higher magnetic field strength ($B$) and elevated proton and electron temperatures, as well as plasma density, compared to other regions. In contrast, while the sheath of ICME2 also shows increased values of magnetic and plasma parameters, these enhancements are less pronounced than those observed for ICME1. This difference may be attributed to the dynamics of the magnetic ejecta (ME) in both ICMEs. For ICME1, the ME exhibits a considerable speed difference relative to the pre-ICME solar wind, resulting in stronger compression and heating of the sheath region. Conversely, ICME2 shows less variation in speed between the pre-ICME solar wind and the ME, leading to a less compressed and heated sheath. Additionally, due to the influence of a high-speed solar wind trailing ICME2, the ME of ICME2 experiences minimal deceleration, maintaining a relatively steady speed. Interestingly, the post-ICME region of ICME2 shows significantly higher values of magnetic field strength and plasma parameters compared to ICME1, suggesting that the ambient medium for the two ICMEs is distinctly different. This difference in the surrounding solar wind environment likely influences the overall evolution of the ICMEs, especially in the later stage of their journey \citep{Manchester2008, Vrsnak2010, Mishra2013}. Notably, the characteristics of the magnetic ejecta for the two ICMEs present an inverse relationship with their speeds. Despite the ME of ICME2 being slower, it exhibits higher values of magnetic field strength and plasma parameters compared to the high-speed ME of ICME1. The derived expansion speeds of the ME regions further highlight this contrast, with ICME1 showing a significantly higher expansion speed of 186 $km$ $s^{-1}$ compared to 35 $km$ $s^{-1}$ for ICME2. This suggests that the ME of ICME1 undergoes stronger expansion dynamics, whereas ICME2’s ME remains more confined. Overall, the differences in the pre-ICME, sheath, ME, and post-ICME regions underscore the diverse physical and dynamical properties of ICME1 and ICME2, influenced by variations in their ambient environments and intrinsic characteristics.

\subsection{Derived thermal states at 1AU}

\begin{figure*}[ht]
    \centering
    \includegraphics[width=0.9\linewidth]{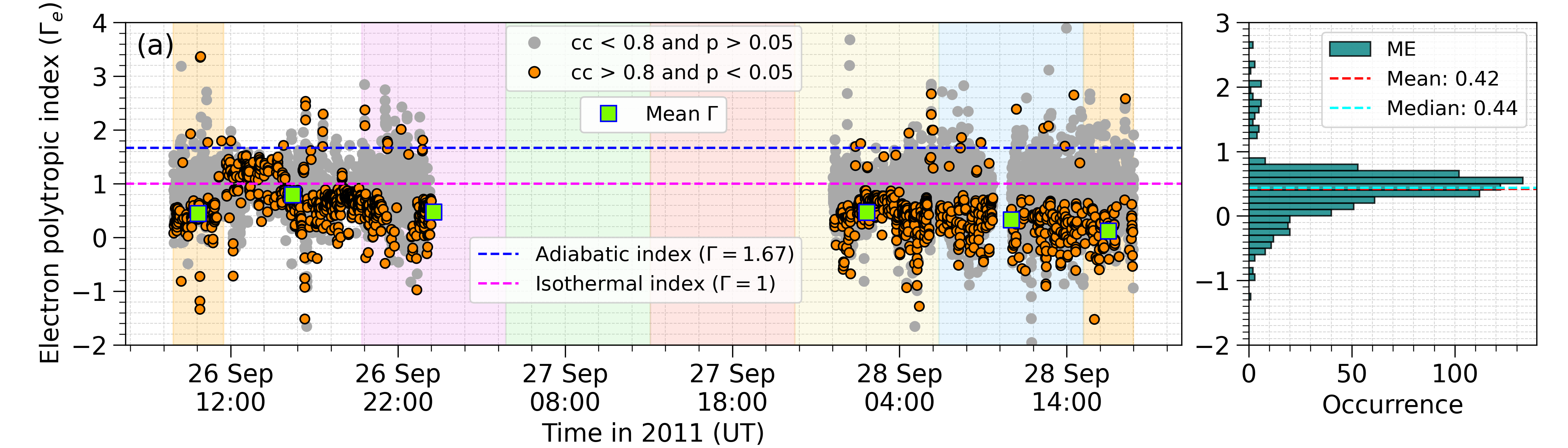}\\
    \includegraphics[width=0.9\linewidth, trim=0 0 0 8, clip]{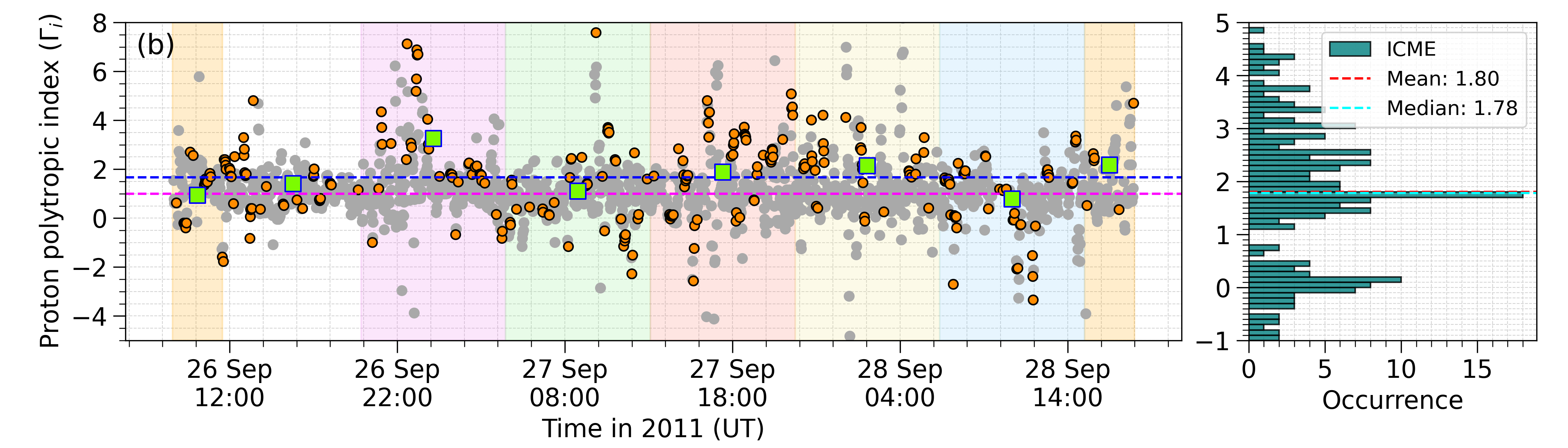}\\
    \includegraphics[width=0.9\linewidth]{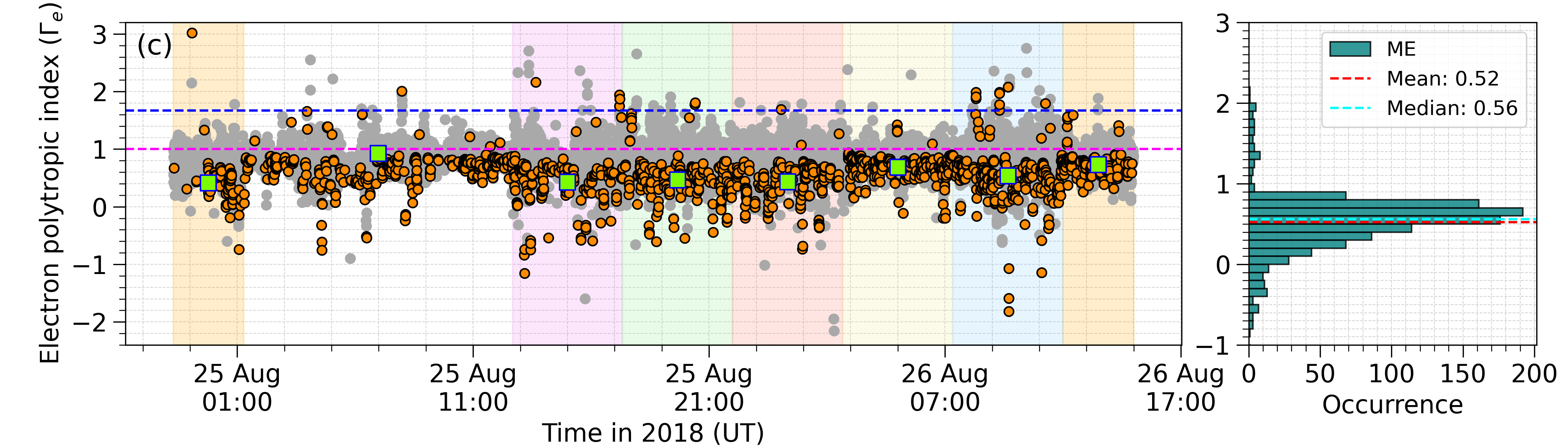}\\
    \includegraphics[width=0.9\linewidth, trim=0 0 0 8, clip]{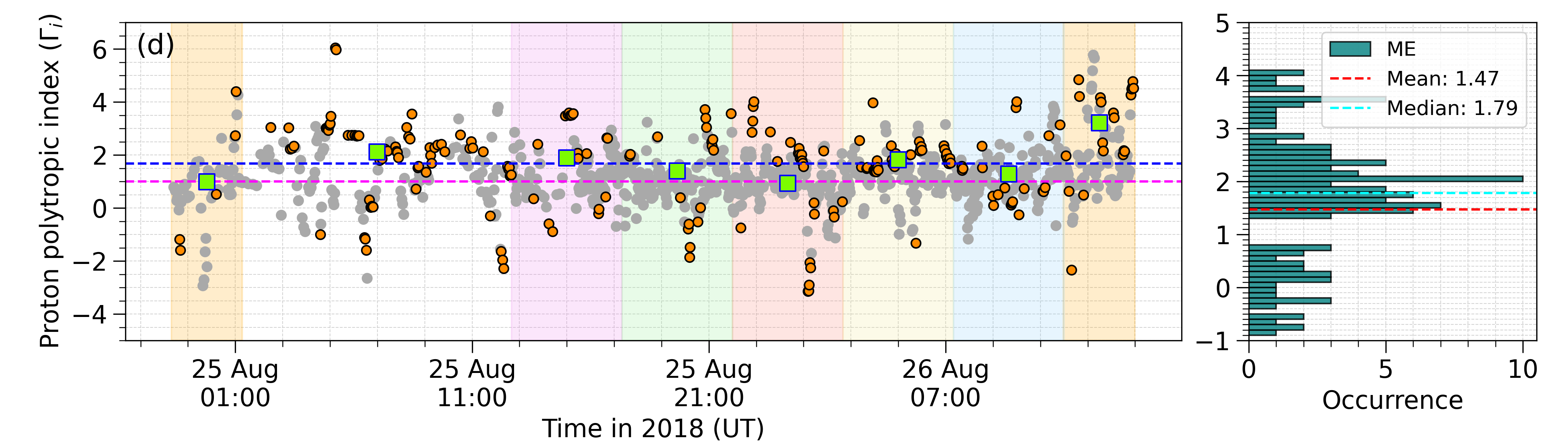}
    \caption{Derived polytropic index ($\Gamma$) values for ICME1 (top two panels) and ICME2 (bottom two panels). (a) and (c) shows the derived electron polytropic index ($\Gamma_e$) using 9-second resolution electron data, whereas (b) and (d) shows the derived proton polytropic index ($\Gamma_p$), using 92-second resolution proton data from Wind/SWE. The orange-shaded regions at the beginning and end represent the pre and post-ICME solar wind. The white-shaded area marks the sheath region, while the multicolored-shaded region between the sheath and post-ICME solar wind corresponds to the magnetic ejecta, divided into five equal parts.}
    \label{fig:insitu_gamma}
\end{figure*}

Figure \ref{fig:insitu_gamma} shows the variations of the polytropic index ($\Gamma$) within pre-ICME, sheath, ME, and post-ICME, the solar wind medium for both electrons and ions. The background shaded color region represents various regions associated with ICME, as discussed in the previous section. In each panel, we overlaid the reference lines for the adiabatic index ($\Gamma =5/3$) and isothermal index ($\Gamma =1$) values. In each panel, we plotted both the reliable (orange color) and unreliable (gray color) $\Gamma$ values obtained from fitting density and temperature. The reliable values correspond to a good fit (CC$>0.8$, and p$<0.5$) of density and temperature to be accepted for further interpretation of the thermal state. Additionally, the mean $\Gamma$ values for each shaded region are overlaid (green squares), and the right side of each panel displays histograms of reliable $\Gamma$ values within the ME duration only.

Figure \ref{fig:insitu_gamma}(a) and (b) show the polytropic index for electron $\Gamma_p$ and proton $\Gamma_e$ in associated structures of ICME1, respectively. In the pre and post-ICME regions, the mean $\Gamma_e$ values are 0.45 and 0.12, respectively, indicating significant electron heating in the ambient medium of ICME1. Both the sheath and ME of ICME1 show a dominant heating state. There is a decrease in the $\Gamma_e$ value from the beginning to the end of the sheath region with a mean $\Gamma_e$ of 0.79, suggesting comparatively less heating than the pre-ICME solar wind. Moving inside the ME of ICME1, the mean $\Gamma_e$ values slowly decrease, indicating more and more heating towards the trailing edge of ICME1. Figure \ref{fig:insitu_wind}(I)(g) further supports this observation with a near-constant electron temperature in Parts 4 and 5 of the ME. Overall, the mean and median $\Gamma_e$ values in the ME are 0.42 and 0.44, respectively, confirming a heating state for electrons within ICME1. The value $\Gamma_e <1$ suggests a negative correlation between electron temperature and density observed in ME has led to differing interpretations in the previous studies. \citet{Osherovich1993poly, Sittler1998} have interpreted this negative correlation as a measure of electron polytropic index in ME plasma. \citet{Hammond1996} argue that the negative correlation in ME results from differing expansion histories of plasma parcels rather than indicating the polytropic index. \citet{Gosling1999} concluded that the negative correlation between electron density and temperature in ME arises from internal structures and the plasma's tendency to maintain local pressure balance

\clearpage

 \noindent as it expands. Thus, in our study, we analyzed the correlation between electron density and temperature within localized subintervals of the ME rather than the entire structure, allowing us to examine plasma parameters in regions with varying polytropic indices.

The thermal states of protons show more variability compared to electrons. The pre and post-ICME1 regions exhibit mean $\Gamma_p$ of 0.93 and 2.17, respectively. This suggests the pre-ICME protons are heated, whereas the post-ICME protons release heat. The sheath of ICME shows a mean $\Gamma_p$ value of 1.4, indicating moderate heating is insufficient to maintain proton temperature. Following the sheath, Part 1 of ME shows a strong heat release state with mean $\Gamma_p$ = 3.2, which can also be noticed as a decrease in the proton temperature profile in Figure \ref{fig:insitu_wind}(I)(e).

All other parts, except Part 5 in ME of ICME1, show a mix of heating and heat release states close to the adiabatic index. The Part 5 has mean $\Gamma_p$ = 0.78, suggesting more heating.  The in-situ observed proton temperature also shows a near-constant value for Part 5 in ME of ICME1 Figure \ref{fig:insitu_wind}(I)(e). Moreover, the mean and median values of the ($\Gamma_p$) distribution inside ME are 1.8 and 1.78, respectively, suggesting an overall heat release consistent with the observed decreasing proton temperature in ICME1 (Figure \ref{fig:insitu_wind}(I)(e)).

Figure \ref{fig:insitu_gamma}(c) and (d) show the polytropic index for electron $\Gamma_e$ and proton $\Gamma_p$, respectively, in ICME2. The pre and post-ICME2 regions show mean $\Gamma_e$ values of 0.42 and 0.73, suggesting dominant electron heating around the ICME2. The mean and median values of the ($\Gamma_e$) distribution inside ME of ICME2 are 0.52 and 0.56, respectively, suggesting strong heating. This enhanced heating is reflected in the higher electron temperatures in the ME compared to the sheath, with a less pronounced decrease (Figure \ref{fig:insitu_wind}(II)(h)).

All the associated structures of ICME2 exhibit mixed values of $\Gamma_p$, suggesting the presence of both heating and heat-releasing states. All parts of ME of ICME2 show a mixture of heating and heat release states close to the adiabatic index. The mean and median values of the ($\Gamma_p$) distribution inside ME of ICME2 are 1.47 and  1.79, respectively. Thus, there will not be much rapid heat release, and there will be less drop in proton temperature across the ME of ICME2 (Figure \ref{fig:insitu_wind}(II)(e)). Interestingly, Part 5 of ME shows a mean $\Gamma_p$ of 1.28; although this suggests a moderate temperature decrease, an increasing trend is observed, possibly due to low proton density in this region. The additional heat is distributed among fewer protons, resulting in a temperature rise. The post-ICME has mean $\Gamma_p$ = 3.2, suggesting rapid heat release, which could be due to the already compressed solar wind because of the trailing CIR and HSS.

Despite differences in the in-situ parameters of the pre and post-ICME regions for ICME1 and ICME2 (Table \ref{tab:cme_table}), the thermal states exhibit similar trends. Both ICMEs show electron heating states in the pre and post-ICME regions. While the pre-ICME regions exhibit moderate proton heating, the post-ICME regions display strong proton heat release in both ICMEs. The ME of ICME1 exhibits stronger electron heating than that of ICME2, while protons show dominant heat release in ICME1 and heating states in ICME2.

\subsection{Connection to the near-Sun thermal states}

\citet{Khuntia2023} analyzed the evolution of the thermal state of CMEs associated with ICME1 and ICME2 with distance from the Sun considering the same polytropic approach but with the help of 3D kinematics and analytical FRIS model. The polytropic index ($\Gamma$) was calculated by considering the whole CME as a single structure with an average temperature, $T=(T_e + T_p )/2$ and a number density, $n=n_e=n_p $. Therefore, an effective polytropic index using both electrons and proton thermal states of the entire ME is needed to be compared with the thermal state of the CME near the Sun reported in \citet{Khuntia2023}. Considering the ($(T_e + T_p)/2$) as the average temperature of ME and $n=n_e=n_p $ (charge neutrality) as the plasma number density, we can write the polytropic equation for the whole ME as \[ T_e + T_p \propto n^{\Gamma_{eff} - 1}\]

Since \(T_e\) is associated with \(\Gamma_e\) and \(T_p\) with \(\Gamma_p\), we can find the weighted average of electron and proton polytropic index. The effective polytropic index \(\Gamma_{eff}\) of the ME as a weighted average of \(\Gamma_e\) and \(\Gamma_p\) can be expressed as
\[\Gamma_{eff} \approx \frac{\Gamma_e T_e + \Gamma_p T_p}{T_e + T_p}\] where the weights are proportional to the temperatures of the electron and proton populations. This approach assumes that the contributions of \(\Gamma_e\) and \(\Gamma_p\) to the overall thermodynamic behavior of the plasma are dictated by the temperature fractions of electrons and protons.

For our study, we can choose the mean electron and proton temperatures of ME as $T_e$ and $T_p$, respectively (Table \ref{tab:cme_table}). The mean value of the $\Gamma$ distribution inside ME can be considered as $\Gamma_e$ and $\Gamma_p$ (Figure \ref{fig:insitu_gamma}). For ICME1, using $T_e = 14.2 \times 10^4 K$ , $T_p = 7.2 \times 10^4 K $ , $\Gamma_e = 0.42 $ and , $\Gamma_p = 1.8$, the effective polytropic index is calculated as $\Gamma_{eff} = 0.88$. Using $T_e = 16.5 \times 10^4 K$ , $T_p = 5.5 \times 10^4 K $ , $\Gamma_e = 0.52 $ and , $\Gamma_p = 1.47$, the effective polytropic index is obtained as $\Gamma_{eff} = 0.76$ for ICME2.  Thus, the ME of both the ICMEs exhibit substantial heating even at 1AU, despite their expansion. Interestingly, a similar near-isothermal state was also derived at heights near the Sun, around 20 $R_\odot$ and 15 $R_\odot$ for the corresponding fast and slow CMEs (Figure 5 and 6 in \citealt{Khuntia2023}), respectively. The near-isothermal state observed at both locations, near the Sun and at 1 AU, indicates substantial deviations from the purely adiabatic expansion expected in interplanetary space, which would typically result in cooling as the ICMEs expand. This consistent heating across large heliospheric distances highlights the importance of continuous and ongoing heating mechanisms in influencing the evolution of ICMEs as they propagate outward from the Sun.

\subsection{Derived turbulence properties at 1AU}

The kinetic processes, such as wave-particle interaction and magnetic reconnection, govern the cross-scale energy transfer in the magnetic plasma, bridging large-scale dynamics and small-scale dissipation. Kinetic processes involve particle-scale dynamics that mediate the transfer or dissipation of energy. These processes often occur in the collisionless regime, where collective electromagnetic fields dominate interactions. The ME or MFRs are large-scale, organized magnetic structures in the heliosphere known for their ability to induce significant disturbances in the magnetosphere \citep{Gonzalez1999, Tsurutani2011, Echer2013}. However, we have a limited understanding of the turbulence characteristics within MFRs \citep{Leamon1998, SorrisoValvo2021, Rodriguez2023, Shaikh2024}. In this study, we analyzed high-resolution magnetic data from the Wind spacecraft to investigate the turbulence properties within two ICMEs, each exhibiting distinct kinematics and interacting with different ambient media.

\begin{figure*}[ht]
    \centering
    \includegraphics[width=0.9\linewidth]{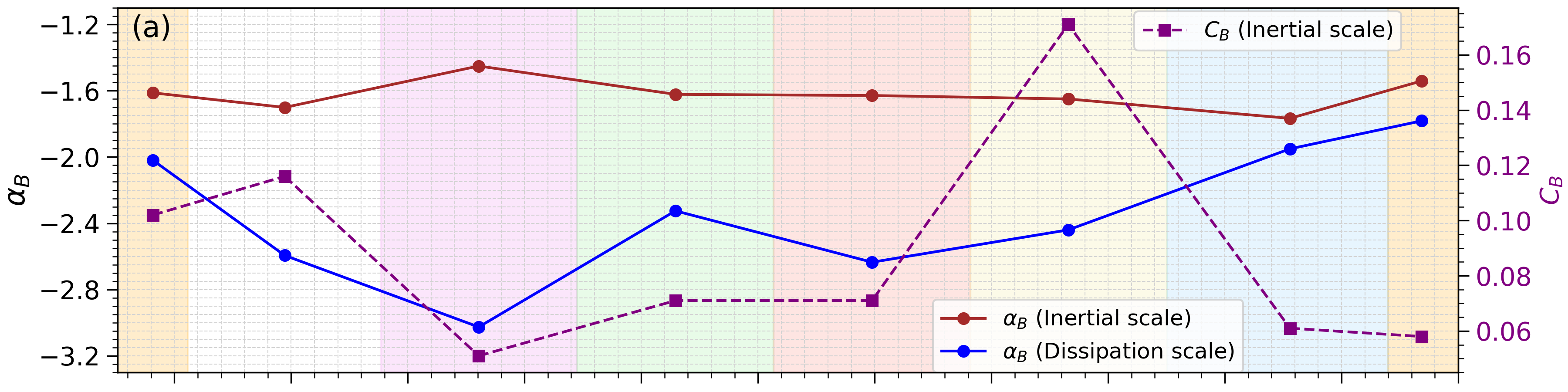}\\ 
    \vspace{-0.1cm}
    \includegraphics[width=0.9\linewidth]{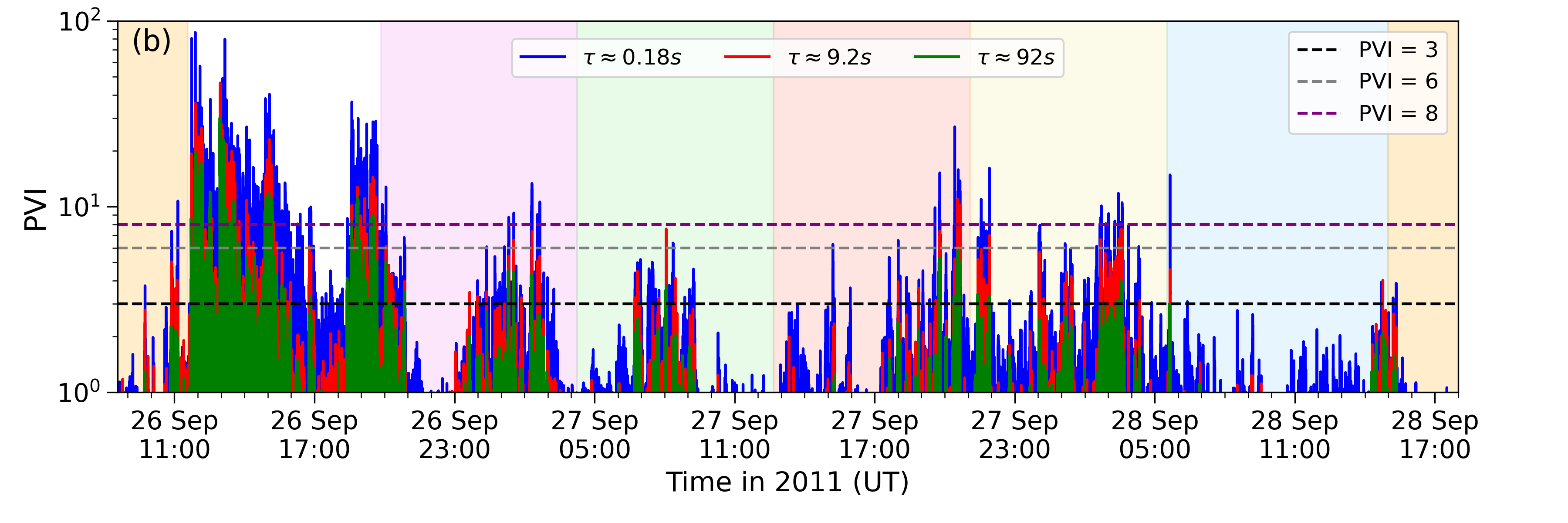}\\
    \includegraphics[width=0.9\linewidth]{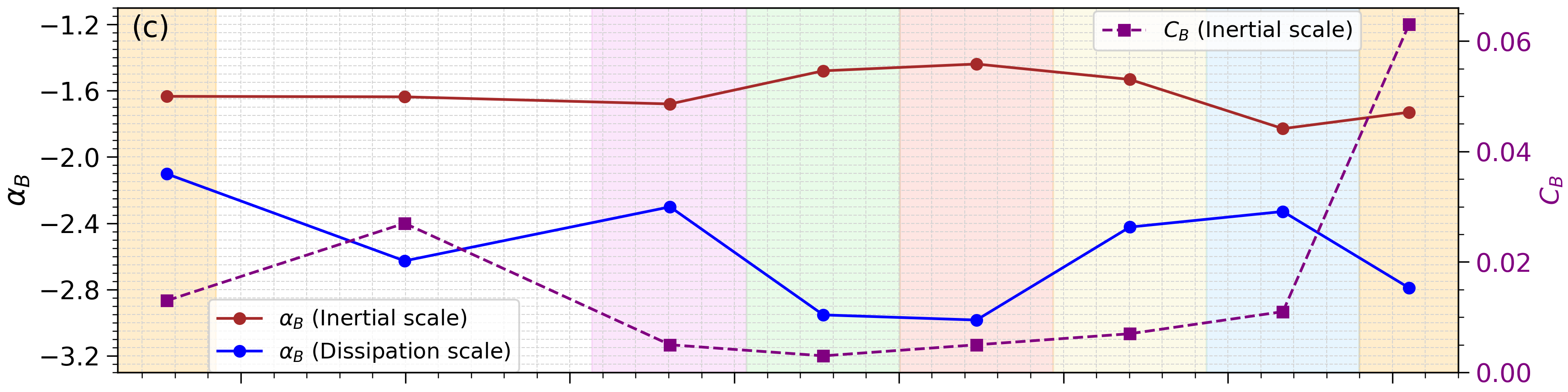}\\
    \vspace{-0.1cm}
    \includegraphics[width=0.9\linewidth]{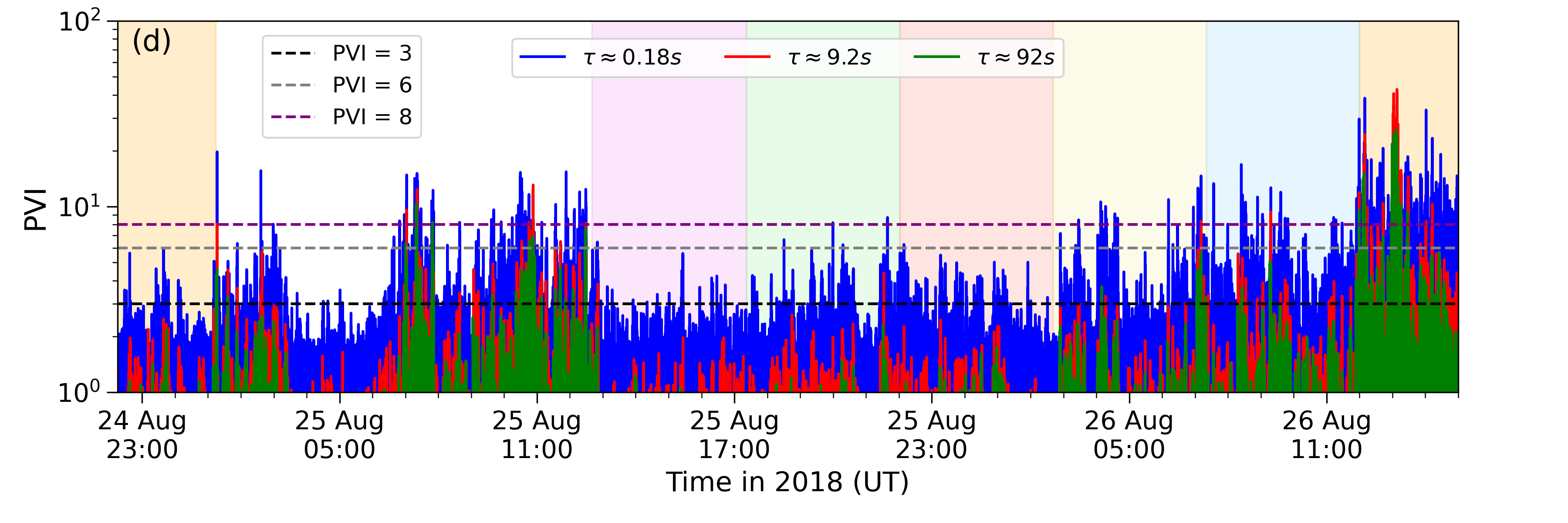}
    \caption{Spectral slope ($\alpha_B$), and magnetic compressibility factor ($C_B$) for ICME1 (a) and ICME2 (c), derived using high-resolution Wind/MFI magnetic data. The derived PVI values for ICME1 (c) and ICME2 (d). The orange-shaded regions at the beginning and end represent the pre and post-ICME solar wind. The white-shaded area marks the sheath region, while the multicolored-shaded region between the sheath and post-ICME solar wind corresponds to the magnetic ejecta, divided into five equal parts. }
    \label{fig:slope_cb_pvi}
\end{figure*}

We did linear fitting to the three-point sliding average of trace power spectra $P_{tr}$ in two frequency regimes, inertia scale ($-3 < \log_{10} f < -0.7 $) and dissipation scale ($-0.3 < \log_{10} f < 0.5 $), to calculate the spectral slope ($\alpha_B$) and magnetic compressibility factors ($C_B$). The fitted profiles for various regions associated with ICME1 and ICME2 are shown in Appendix (Figures \ref{fig:fast_five_me}, \ref{fig:slow_five_me}, \ref{fig:fast_spec_pre_post},  \ref{fig:slow_spec_pre_post}). Figure \ref{fig:slope_cb_pvi} shows the derived $\alpha_B$, $C_B$, and PVI values for each color-shaded region of both ICMEs. The shaded background color represents different regions of the ICME, as described in the previous section. Figure \ref{fig:slope_cb_pvi}(a) and (c) shows $\alpha_B$ for both inertial scales (brown-dot line) and dissipation scales (blue-dot line) along with the derived average $C_B$ values in the inertial scale of each region.

In the inertial scale, the spectral slope $\alpha_B$ is around  $-1.6$ for the pre-ICME solar wind, whereas it is around $-1.7$ for the sheath region of ICME1 (Figure \ref{fig:slope_cb_pvi}a). Thus, standard Kolmogorov \citep{Kolmogorov1941} fully developed turbulence is present in both pre-ICME and sheath regions of ICME1. A steeper $\alpha_B$ suggests the energy is more efficiently transferred to smaller scales through turbulent processes. So, the sheath has a stronger turbulence where the energy transfer rate is higher. Except for Part 1 in ME, other parts show $\alpha_B$ close to  $-1.6$ for the inertial range, which is consistent with ICME magnetic flux rope \citep{Borovsky2019, Good2023} turbulence spectra at 1 au. Part 1 of ME and the post-ICME regions show a shallower spectrum with $\alpha_B$ close to $-1.5$ for the inertial range, suggesting that the nature of turbulence is close to the Iroshnikov–Kraichnan (IK) spectrum \citep{Iroshnikov1963, Kraichnan1965}, indicating underdeveloped turbulence \citep{Bruno2013}.

In the dissipation scale, turbulence cascades terminate, and energy dissipates into heat.  The spectral slope $\alpha_B$ decreases from the pre-ICME solar wind, then sheath, and up to Part 1 in ME of ICME1 (Figure \ref{fig:slope_cb_pvi}(a)). A more steep $\alpha_B$ in the dissipation scale suggests a higher dissipation at a smaller scale. Thus, the dissipation mechanism is more efficient for Part 1 in ME, followed by the sheath and pre-ICME region. Further,  $\alpha_B$ decreases up to the post-ICME regions of ICME1, suggesting less and less efficient heat dissipation at the later parts of ME of ICME1. This is also evident from the decreasing proton and electron temperature profile from the in-situ measurements (Figure \ref{fig:insitu_wind}(I)(e and g)).

The compressibility factor, defined as $C_B=P_t/P_{tr}$, i.e., the ratio between the power spectral density of the magnetic field magnitude and vector fluctuations \citep{Bavassano1982}, provides information about the Alfvénic nature of the plasma. Low magnetic compressibility ($C_B <<1$) indicates that the fluctuations do not significantly change the magnitude of the magnetic field but involve directional changes, characteristic of non-compressive Alfvénic fluctuations. $C_B <<1$ within certain regions would indicate the dominance of Alfvénic fluctuations over compressive fluctuations, suggesting a relatively high Alfvénic content \citep{Bruno1993, Damicis2015}. The pre-ICME has a higher value of $C_B$ than the post-ICME region of ICME1 (Figure \ref{fig:slope_cb_pvi}(a)). Part 4 in ME of ICME1 shows a higher $C_B$ than other ME regions, suggesting the presence of comparatively more compressive fluctuations. Overall, the ME of ICME1 has a lower $C_B$ value than the sheath and pre-ICME regions, indicating the presence of high Alfvénic fluctuations \citep{Bruno2016, Damicis2021, Telloni2021}.

\clearpage

In the inertial scale, the spectral index $\alpha_B$ is around  $-1.6$ for the pre-ICME solar wind and sheath of ICME2, suggesting the presence of standard Kolmogorov fully developed turbulence (Figure \ref{fig:slope_cb_pvi}(c)). The spectral index $\alpha_B$ is shallower towards the central part in ME compared to the leading and trailing edge, indicating less developed turbulence. It implies that the energy is not cascading efficiently from injection to dissipation scale towards the central part of ME of ICME2. In the dissipation scale, the central part in ME has a steeper $\alpha_B$ than the outer ones, indicating stronger dissipation of energy (Figure \ref{fig:slope_cb_pvi}(c)).  \citet{Riazantseva2019} divided ICMEs into two parts, the magnetic cloud (MC) and the ME, and found that the MC exhibits a shallower spectral slope at the inertial scale and a steeper spectral slope at the dissipation scale compared to the ME, which is consistent with our findings for ICME1 (ME) and ICME2 (MC).

Furthermore, $C_B$ is less in ME compared to the surrounding sheath and post-ICME regions. The high  $C_B$ value of the post-ICME region indicates the dominance of compressive structures, such as slow magnetosonic waves, shocks, and density enhancements, which can efficiently transfer energy. This is also evident from the steeper $\alpha_B$ in the dissipation scale for the post-ICME region of ICME2. These variations in spectral behavior between the central and outer regions highlight the spatially varying turbulence dynamics and dissipation processes within the ME, influenced by the local plasma conditions and magnetic field structure.

Temperature anisotropies in protons can excite kinetic instabilities like the proton cyclotron or mirror instabilities, modifying the dissipation processes \citep{Gary1993, Denton1994, Verscharen2019}. If anisotropies are stronger, dissipation may be stronger, leading to a steeper $\alpha_B$ in dissipation scale. The higher proton temperature anisotropy in the sheath compared to other regions of ICME1 corresponds to the steeper $\alpha_B$ in the dissipation scale. The sheath and ME of ICME2 have comparable moderate proton temperature anisotropy. However, the steeper $\alpha_B$ in the dissipation scale for the ME of ICME2 could be due to the lower $C_B$ value. The proton and electron anisotropy is less in part 1 of ME of ICME1 (Figure \ref{fig:insitu_wind}(I)(e and g)); however, we still notice stronger dissipation (Figure \ref{fig:slope_cb_pvi}a). In ME of ICME2, Part 1 exhibits lower $T_\perp$, while Part 5 shows lower $T_\parallel$ for proton (Figure \ref{fig:insitu_wind}(II)(e)), yet both regions share a shallow spectral index ($\alpha_B$) in the dissipation scale and low magnetic compressibility (Figure \ref{fig:slope_cb_pvi}(c)). In Part 1 of ICME2, the reduced $T_\perp$ could imply weaker perpendicular heating or dissipation mechanisms like cyclotron damping or the presence of whistler waves \citep{Hellinger2000}, while in Part 5 of ICME2, the lower $T_\parallel$ hints at less effective parallel heating through processes like Landau damping or more effective mirror mode waves \citep{AlaLahti2019}. This contrast suggests that distinct dissipation mechanisms dominate in each region. Overall, the ME of ICME1 has stronger proton and electron temperature anisotropy with a higher $C_B$ and less steeper $\alpha_B$ in dissipation scale compared to the ME of ICME2.

\subsection{Measured Intermittencies at 1AU}

In this study, we also employ the Partial Variance of Increments (PVI) method to detect various intermittencies. The PVI is a reliable measure for detecting sharp gradients in the magnetic field, such as discontinuities, current sheets, and sites of magnetic reconnection \citep{Greco2018}. These discontinuities are thought to be crucial for enhanced dissipation, as well as particle heating and energization in space plasmas \citep{Tessein2013, Chasapis2015}. We examine how the value of PVI varies with respect to the spectral slope ($\alpha_B$) in both the inertial and dissipation scale and also with the compressibility factor ($C_B$). Figure \ref{fig:slope_cb_pvi}(b) and (d) show the magnetic PVI values for various regions associated with ICME1 and ICME2, respectively. We have calculated PVI for three different time lags ($\tau$), such as 0.18s, 9.2s, and 92s, to include both inertial and dissipation scales. We overlaid the three threshold PVI values at 3, 6, and 8. The transition from Gaussian to non-Gaussian magnetic field fluctuations is observed around PVI = 3 \citep{Osman2011a, Osman2011b}. Higher PVI values primarily correspond to the non-Gaussian heavy tails of the magnetic field increment distribution.

All three time lags $\tau$ chosen for our PVI analysis identify a significant number of non-Gaussian events (PVI $>$ 3), but the smaller $\tau$ detects more current sheets (PVI $>$ 6), and potential reconnection sites (PVI $>$ 8)\citep{Rohit2020}. Notably, the sheath, Part 3, and Part 4 in ME of ICME1 exhibit higher PVI values compared to other regions of ICME1 (Figure \ref{fig:slope_cb_pvi}(b)). Similarly, in ICME2, the sheath, Part 4 and Part 5 of the ME, as well as the post-ICME solar wind, show increased PVI values compared to other structures of ICME2 (Figure \ref{fig:slope_cb_pvi}(d)). The sheath regions of both ICMEs exhibit enhanced intermittency and compressibility compared to the ME, aligning well with the findings of case studies by \citet{Kilpua2020} and \citet{Good2020}. We also observed fewer intermittencies in the pre-ICME solar wind, which supports the statistical findings of \citep{Kilpua2021}.

We found higher PVI values in a region corresponding to a higher value of $C_B$. Thus, there is a good qualitative correlation between $C_B$ and the PVI values for both ICME1 and ICME2, indicating a relationship between magnetic field fluctuations and plasma compressibility. $C_B$ measures the extent to which the magnetic field changes align with density fluctuations, indicating the degree of compressive fluctuations in the plasma. Higher PVI values represent regions with sharp magnetic field gradients, often coinciding with intense localized activity. These regions will likely have more $C_B$, as the dynamic interactions and energy dissipation processes in such areas can compress the plasma more effectively. This relationship implies that areas with strong intermittency (high PVI) are also associated with significant compressive effects, emphasizing the interconnected nature of turbulence and compressibility in ICMEs.

As shown in Figure \ref{fig:slope_cb_pvi}(b) and (d), the PVI measure captures highly intermittent events at smaller $\tau$, whereas such events are less frequent in the time series computed with a 92 s lag. This suggests that smaller lag intervals are more effective in identifying fine-scale structures and dynamic processes in turbulent plasmas, while larger intervals provide a more averaged view of the system. \citet{Osman2011a} utilized the PVI method to associate solar wind discontinuities with localized plasma heating, demonstrating that higher PVI values correlate with increased electron heat flux, electron temperature, and proton temperature. Further analysis by \citet{Osman2011b} showed that the average PVI statistic displays distinct patterns when conditioned on parallel plasma beta and proton temperature anisotropy, with the highest PVI values observed in regions where these parameters reach extreme levels. In our analysis, this correlation holds true in the sheath regions of both ICMEs, where elevated temperature anisotropy and PVI values are evident. However, this pattern is less pronounced within the magnetic ejecta (ME) of both ICMEs.

Comparing the sheaths of both ICMEs, we observed more intermittencies in ICME1 than in ICME2 (Figure \ref{fig:slope_cb_pvi} (b) and (d)), along with stronger mean heating states (Figure \ref{fig:insitu_gamma}). Notably, within the ME of ICME1, the regions with higher PVI values correlate well with $\Gamma_p$, suggesting that increased intermittencies contribute to greater heat release. However, this correlation is not evident in the ME of ICME2. Overall, the results indicate that turbulence and dissipation vary significantly across ICME regions and ambient pre and post-ICME regions, with the sheath showing stronger turbulence and dissipation than the ME. The correlation between PVI values and $C_B$ highlights the interplay between localized intermittent structures and compressive fluctuations, emphasizing the role of compressive effects in energy transfer and dissipation within ICMEs and solar wind.

\section{Conclusions}

This study investigated the thermal and turbulence characteristics of two distinct ICME events, with one exhibiting a fast speed (ICME1) and the other slow speed (ICME2), as observed by the Wind spacecraft at 1 AU. Our primary goal was to understand the variations in the thermal states and turbulence properties within ICME plasma, as well as to examine how the solar wind environment influences these characteristics. Notable findings from this study are summarized below,

\begin{enumerate}
    \item The faster ICME1 exhibited more stronger electron heating compared to the slower ICME2, particularly in the overall ME and sheath regions. Conversely, ICME2 exhibited stronger proton heating within the ME than ICME1. In ICME1, both electron and proton temperatures gradually decreased throughout the ME, whereas ICME2 maintained a nearly constant temperature in the leading parts of the ME, followed by a temperature decrease in the trailing parts.

    \item The near-isothermal state observed near the Sun (15-20 $R_\odot$) and at 1 AU suggests the presence of continuous and efficient heating mechanisms influencing ICME evolution, despite their expansion.

    \item At the inertial scale, all associated structures of ICME1 exhibit near-standard Kolmogorov turbulence with a spectral slope of $\alpha_B \approx -5/3$, whereas the central regions of the ME in ICME2 show a shallower $\alpha_B$, indicating less developed turbulence. At the dissipation scale, ICME1 shows a decreasing trend in $\alpha_B$ from the leading to trailing parts of the ME, while ICME2 exhibits stronger dissipation in the central regions of the ME compared to the outer parts. 
    
    \item The variation of $\alpha_B$ in dissipation scale also indicates that the ME of slower ICME2 is less affected by the ambient medium than the faster ICME2.

    \item Higher proton temperature anisotropy in the sheath region correlates with stronger dissipation, suggesting that proton temperature anisotropies can trigger kinetic instabilities, enhancing dissipation and steepening $\alpha_B$ at the dissipation scale.

    \item PVI values correlate with enhanced plasma heating, with higher PVI values corresponding to increased electron and proton temperature, particularly in sheath regions of both ICMEs.
    
    \item Higher PVI values are found in regions with higher $C_B$, suggesting a qualitative correlation between magnetic field fluctuations, plasma compressibility, and intermittency. Higher PVI regions are associated with stronger compressive effects and localized activity.

    \item The thermal states of the ambient medium (pre and post-ICME solar wind) are similar for both ICMEs, but their turbulence properties differ. ICME1's ambient medium has minimal intermittencies, while ICME2's post-ICME region shows significant intermittencies and stronger turbulent dissipation.

\end{enumerate}

This study highlights the complex thermal and turbulence nature of ICMEs, where variations in the ambient solar wind and the magnetic structure of the ICME itself influence the plasma properties. The differences between ICME1 and ICME2 underscore the importance of understanding these variations to better predict the evolution of ICMEs.


\appendix

\section{Turbulence Characteristics Across Various Regions of ICME1 and ICME2}

\begin{figure*}[ht]
    \centering
    \includegraphics[width=0.9\linewidth]{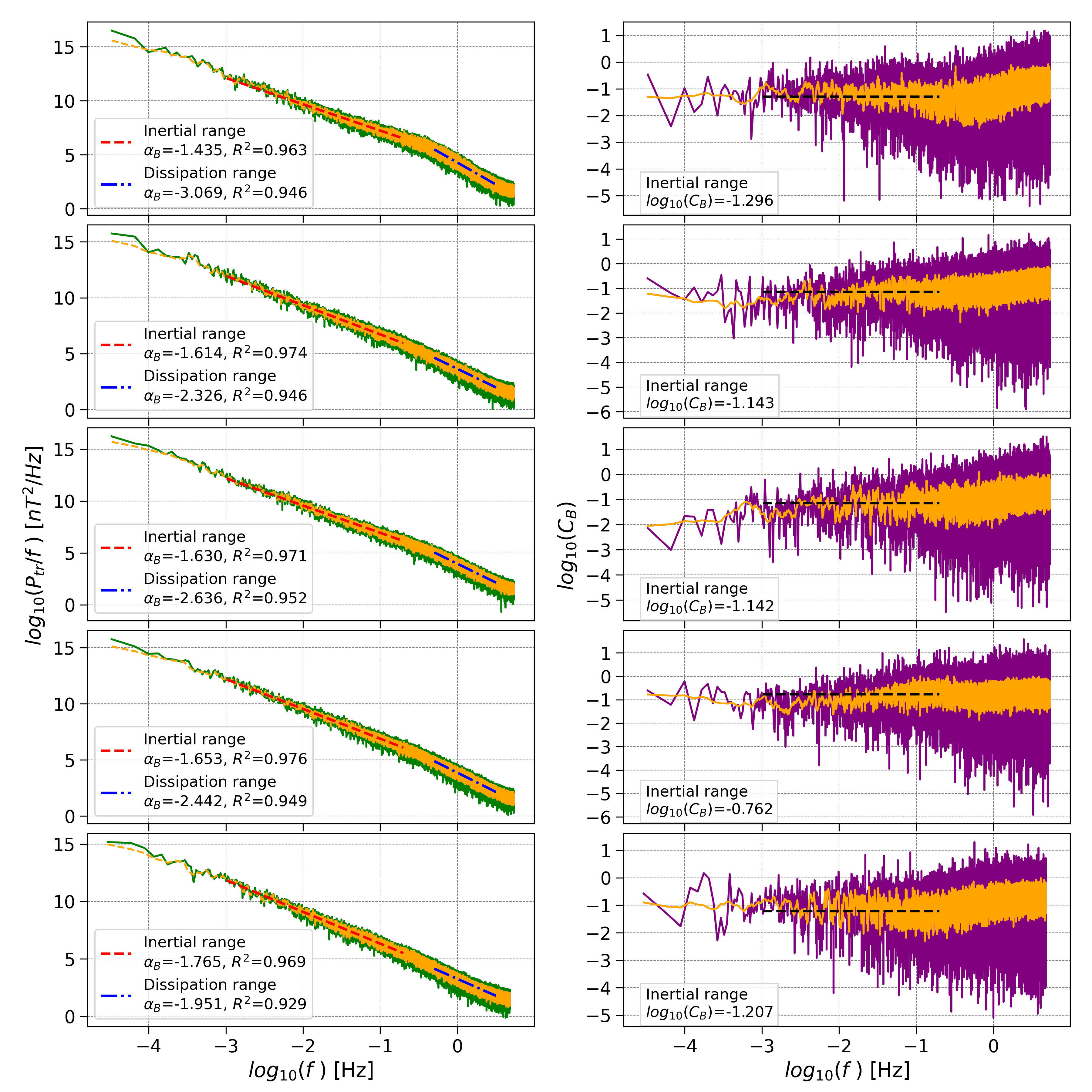}
    \caption{ Left panels: variation of trace power spectra ($P_{tr}$) per frequency across five parts in ME of ICME1. The yellow curve represents the 3-point sliding average of $P_{tr}$. We have overplotted the linear fitted dash lines (red and blue) for both inertial ($-3 < \log_{10} f < -0.7 $) and dissipation ($-0.3 < \log_{10} f < 0.5 $) scales along with the corresponding spectral slope ($\alpha_B$) and goodness of fit ($R^2$) are mentioned. Right panels: variation of magnetic compressibility factor ($C_B$) with frequency across five parts in ME of ICME1.  The yellow curve represents 10-point sliding average values of $C_B$. We have overlaid a black dashed line showing the average value of $C_B$ over the inertial scale.}
    \label{fig:fast_five_me}
\end{figure*}

\begin{figure*}[ht]
    \centering
    \includegraphics[width=\linewidth]{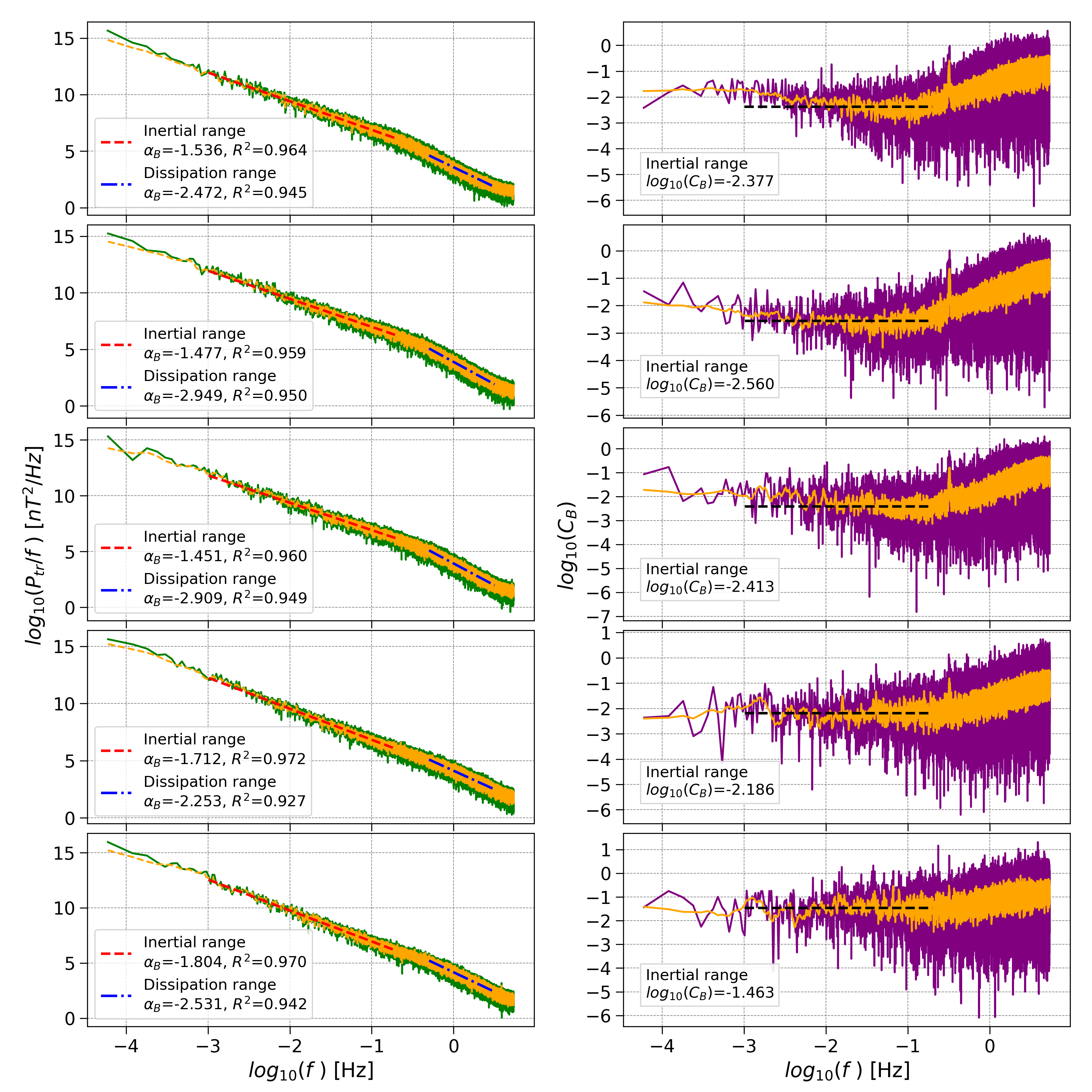}
    \caption{ Left panels: variation of trace power spectra ($P_{tr}$) per frequency across five parts in ME of ICME2. The yellow curve represents the 3-point sliding average of $P_{tr}$. We have overplotted the linear fitted dash lines (red and blue) for both inertial ($-3 < \log_{10} f < -0.7 $) and dissipation ($-0.3 < \log_{10} f < 0.5 $) scales along with the corresponding spectral slope ($\alpha_B$) and goodness of fit ($R^2$) are mentioned. Right panels: variation of magnetic compressibility factor ($C_B$) with frequency across five parts in ME of ICME2.  The yellow curve represents 10-point sliding average values of $C_B$. We have overlaid a black dashed line showing the average value of $C_B$ over the inertial scale.}
    \label{fig:slow_five_me}
\end{figure*}

\begin{figure*}[ht]
    \centering
    \includegraphics[width=0.3\linewidth]{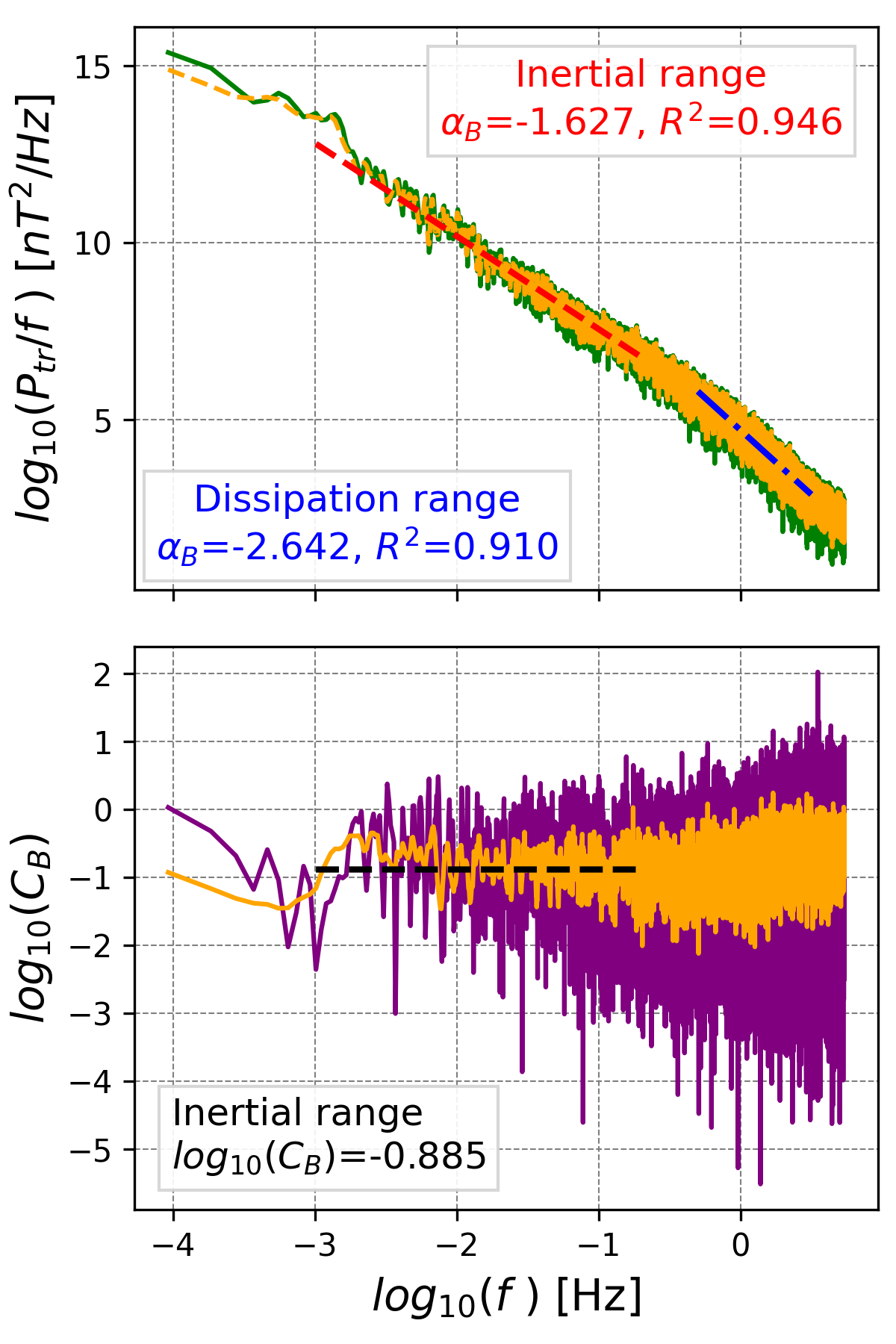}
    \includegraphics[width=0.3\linewidth]{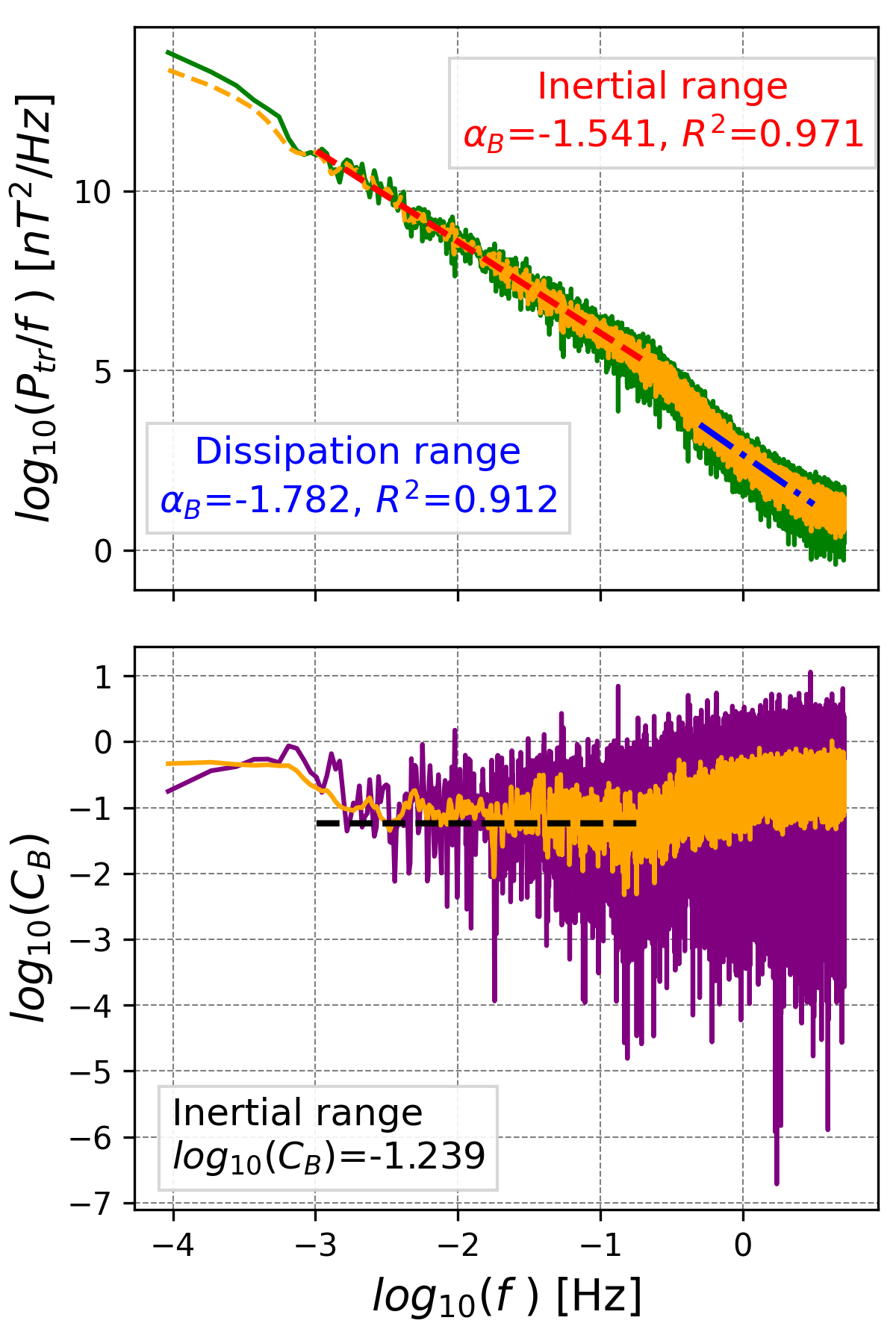}
    \caption{ Variation of trace power spectra ($P_{tr}$) per frequency (top panels) and magnetic compressibility factor ($C_B$) with frequency (bottom panels) across pre-ICME (left panels) and post-ICME (right panels) regions of ICME1. The top panels include a 3-point sliding average (yellow curve) and linear fits (dashed red and blue lines) for the inertial ($-3 < \log_{10} f < -0.7$) and dissipation ($-0.3 < \log_{10} f < 0.5$) scales, with the spectral slope ($\alpha_B$) and goodness of fit ($R^2$) provided. The bottom panels show 10-point sliding averages of $C_B$ (yellow curve), with a black dashed line indicating the average $C_B$ over the inertial scale.}
    \label{fig:fast_spec_pre_post}
\end{figure*}

\begin{figure*}[ht]
    \centering
    \includegraphics[width=0.3\linewidth]{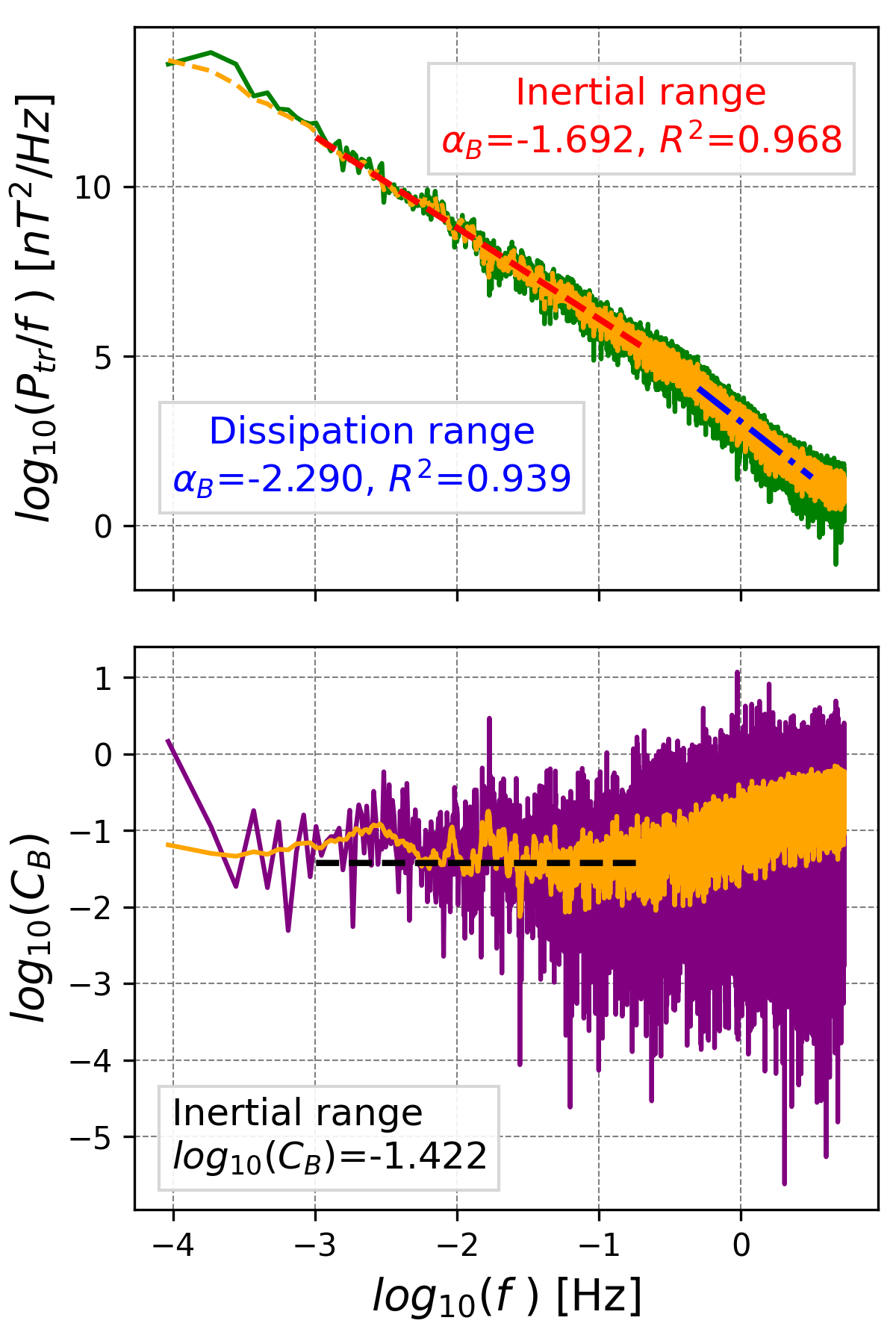}
    \includegraphics[width=0.3\linewidth]{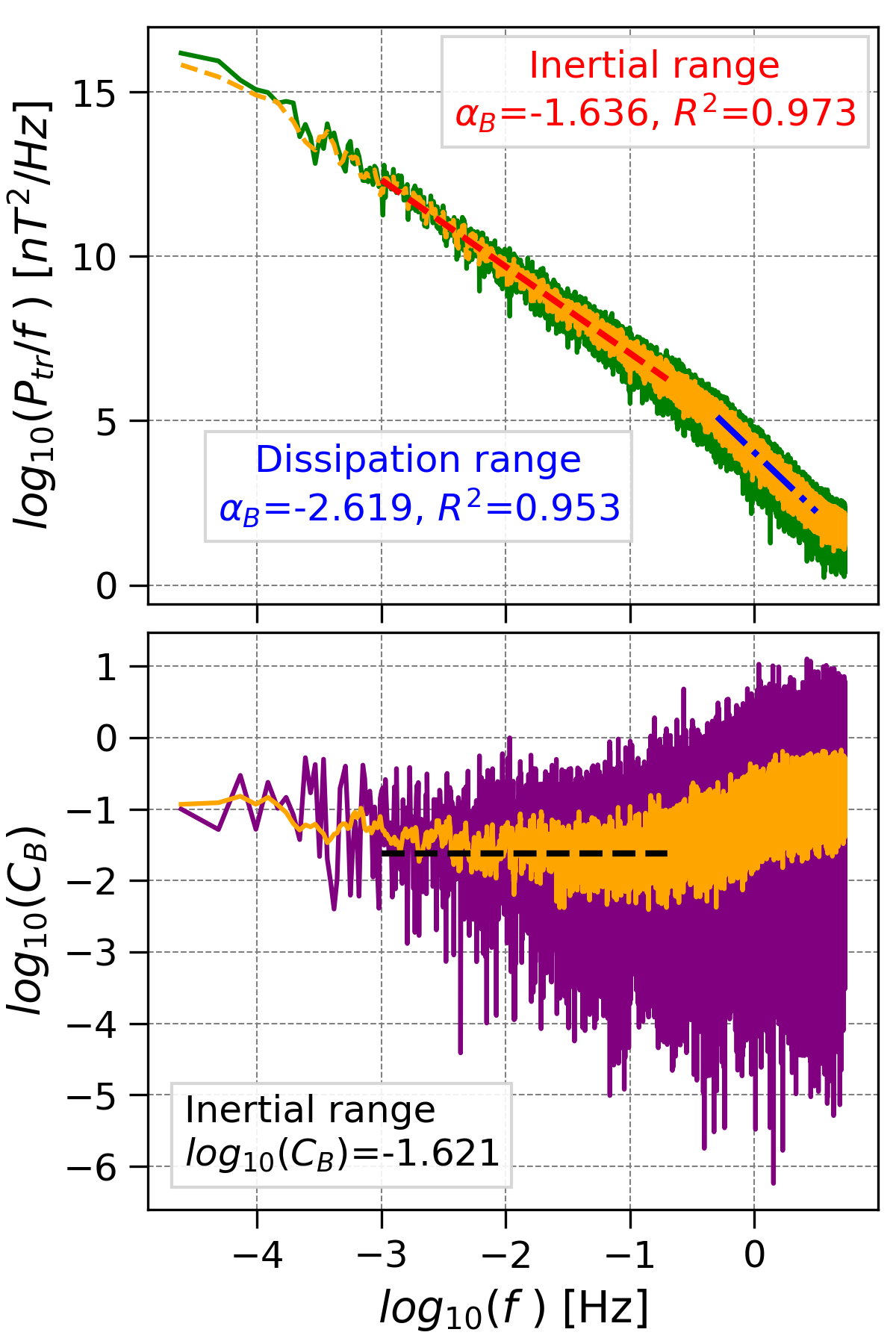}
    \includegraphics[width=0.3\linewidth]{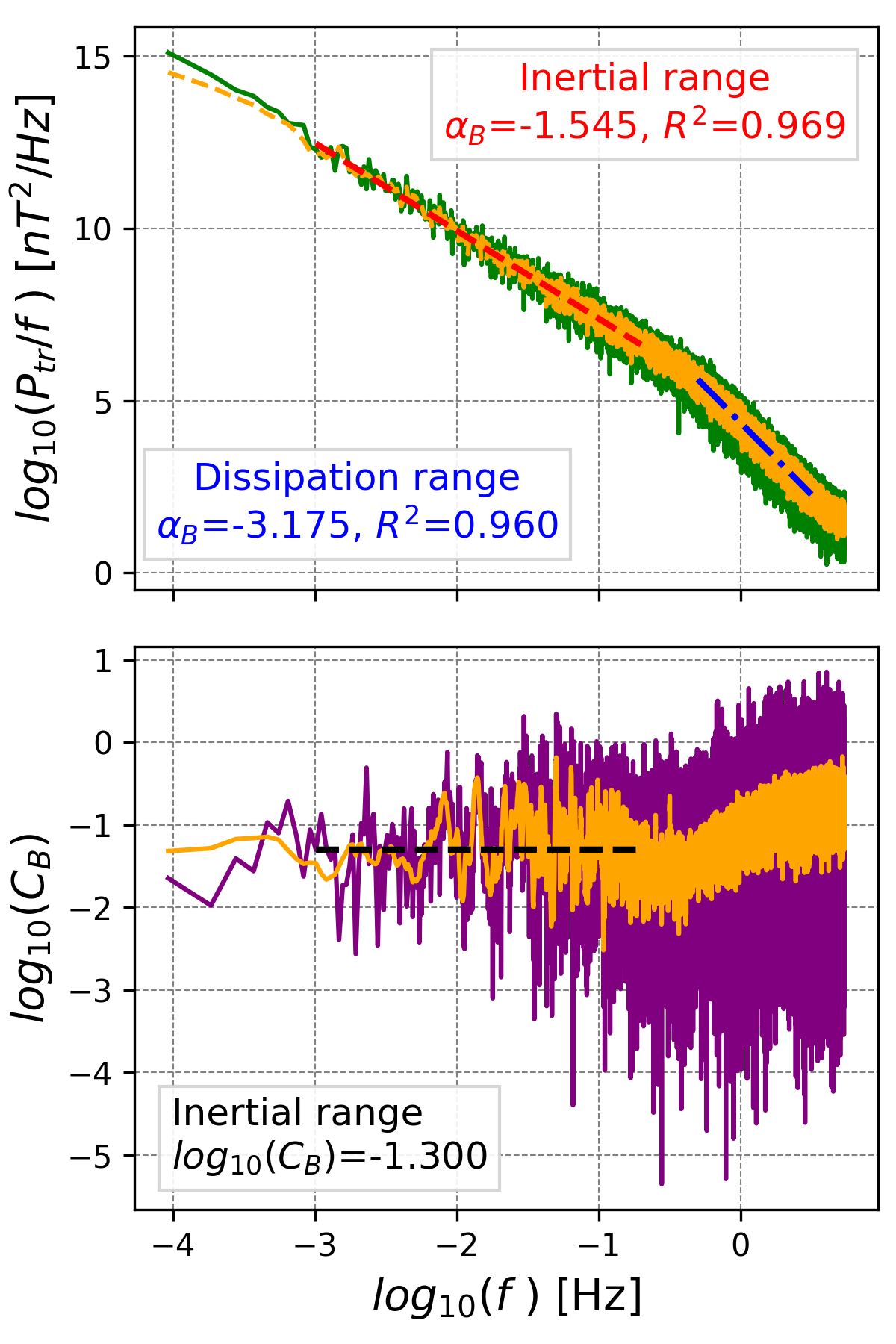}
    \caption{  Variation of trace power spectra ($P_{tr}$) per frequency (top panels) and magnetic compressibility factor ($C_B$) with frequency (bottom panels) across pre-ICME (left panels), sheath (middle panels) and post-ICME (right panels) regions of ICME2. The top panels include a 3-point sliding average (yellow curve) and linear fits (dashed red and blue lines) for the inertial ($-3 < \log_{10} f < -0.7$) and dissipation ($-0.3 < \log_{10} f < 0.5$) scales, with the spectral slope ($\alpha_B$) and goodness of fit ($R^2$) provided. The bottom panels show 10-point sliding averages of $C_B$ (yellow curve), with a black dashed line indicating the average $C_B$ over the inertial scale. }
    \label{fig:slow_spec_pre_post}
\end{figure*}

\section*{Acknowledgements}

We acknowledge the instruments team members for providing the Wind spacecraft data, particularly MFI and SWE instrument data. All the observational input data sets used in this study are publicly accessible through the CDAWeb repository (\url{https://cdaweb.gsfc.nasa.gov/}). We would like to express our appreciation to the anonymous referee for their constructive and insightful feedback, which greatly helped to enhance the clarity and quality of this manuscript.

\end{document}